\documentclass[12pt]{article}

\def\be{\begin{equation}}
\def\ee{\end{equation}}
\def\bea{\begin{eqnarray}}
\def\eea{\end{eqnarray}}

\usepackage{epsfig}

\usepackage{amssymb,latexsym}
\usepackage{graphicx}
\usepackage{amssymb, amsmath, amsopn, amsthm}
\usepackage[section]{placeins}

\makeatletter \@addtoreset{equation}{section}
\makeatother

\begin{document}
\thispagestyle{empty}
\hskip 1 cm
\vskip 0.5cm

\vspace{25pt}
\begin{center}
    { \LARGE{\bf Inflationary Cosmology}}\footnote{
Based on a talk given at the 22nd IAP Colloquium, ÒInflation+25Ó, Paris, June 
2006}
    \vspace{33pt}

  {\large  {\bf  Andrei Linde}}

    \vspace{10pt}

    \vspace{10pt} {Department of Physics,
    Stanford University, Stanford, CA 94305}

    \vspace{20pt}
 \end{center}

\begin{abstract}

I give a general review of the history of inflationary cosmology and of its present status.\end{abstract}

\

\tableofcontents

\newpage

\section {Brief history of inflation}

Since inflationary theory is now more than 25 years old, perhaps it is not inappropriate to start this paper with a brief history of the first stages of its development.

Several ingredients of inflationary cosmology were discovered in the beginning of the 70's. The first realization was that the energy density of a scalar field plays the role of the vacuum energy/cosmological constant \cite{Linde:1974at}, which was changing during the cosmological phase transitions \cite{Kirzhnits:1972ut}. In certain cases these changes  occur discontinuously, due to first order phase transitions from a supercooled vacuum state (false vacuum) \cite{Kirzhnits:1976ts}.

In 1978, Gennady Chibisov and I tried to use these facts to construct a cosmological model involving exponential expansion of the universe in the supercooled vacuum as a source of the entropy of the universe, but we immediately realized that the universe becomes very inhomogeneous after the bubble wall collisions. I mentioned our work in my review article \cite{Linde:1978px}, but did not pursue this idea any further.

The first semi-realistic model of inflationary type was proposed by Alexei
Starobinsky in 1979 - 1980 \cite{Star}. It was based on investigation of a
conformal anomaly in quantum gravity.  His model was rather complicated, and its goal was somewhat different from the goals of inflationary cosmology. Instead of attempting to solve the homogeneity and isotropy problems, Starobinsky considered the model of  the universe which was homogeneous and isotropic from the very beginning, and emphasized that his scenario was ``the extreme opposite of Misner's initial chaos.'' 

On the other hand, Starobinsky's model did not suffer from the graceful exit problem, and it was the first model predicting  gravitational waves with a flat spectrum \cite{Star}. The first mechanism of production of adiabatic perturbations of the metric with a flat spectrum, which are responsible for galaxy production, and which were found by the observations of the CMB anisotropy,  was proposed by  Mukhanov  and Chibisov \cite{Mukh} in the context of this model.

A much simpler inflationary model with a very clear physical
motivation was proposed by Alan Guth in 1981 \cite{Guth}.  His model,
which is now called ``old inflation,'' was based on the theory of
supercooling during the cosmological phase transitions
\cite{Kirzhnits:1976ts}. Even though this scenario did not work,  it
played a profound role in the development of inflationary
cosmology since it contained a very clear explanation how
inflation may solve the major cosmological problems.

According to this scenario,  inflation is as   exponential
expansion of the universe in a supercooled false vacuum state.
False vacuum is a metastable state without any fields or particles
but with large energy density. Imagine a universe filled with such
``heavy nothing.'' When the universe expands, empty space remains
empty, so its energy density does not change. The universe with a
constant energy density expands exponentially, thus we have
inflation in the false vacuum. This expansion makes the universe very big and very flat. Then the false vacuum decays, the
bubbles of the new phase collide, and our universe becomes hot.

Unfortunately, this simple and intuitive picture of inflation in the false vacuum state is somewhat misleading. If the probability of the bubble formation is large, bubbles of the new phase are formed near each other, inflation is too short to solve any problems, and the bubble wall collisions make the universe extremely inhomogeneous. If they are formed far away from each other, which is the case if the probability of their formation is small and inflation is long, each of these bubbles represents a separate open universe with a vanishingly small $\Omega$. Both options are unacceptable, which has lead to the conclusion that this scenario does not work and cannot be improved (graceful exit problem) \cite{Guth,Hawking:1982ga,Guth:1982pn}.

The solution was found in 1981 - 1982 with the invention of the new inflationary theory \cite{New}, see also \cite{New2}. In this theory, inflation may
begin either in the false vacuum,  or in an unstable state at the
top of the effective potential. Then the inflaton field $\phi$
slowly rolls down to the minimum of its effective potential.  The
motion of the field away from the false vacuum is of crucial
importance: density perturbations produced during the slow-roll
inflation are inversely proportional to $\dot \phi$
\cite{Mukh,Hawk,Mukh2}. Thus the key difference between the new
inflationary scenario and the old one is that the useful part of
inflation in the new scenario, which is responsible for the
homogeneity of our universe, does {\it not} occur in the false
vacuum state, where $\dot\phi =0$.

Soon after the invention of the new inflationary scenario it became so popular that even now most of the textbooks on astrophysics incorrectly describe inflation as an exponential expansion in a supercooled false vacuum state during the cosmological phase transitions in grand unified theories. Unfortunately, this scenario was plagued by its own problems. It works only if the effective potential of the field $\phi$ has a very a flat plateau near $\phi = 0$, which is somewhat artificial. In most versions of this
scenario the inflaton field has an extremely small coupling
constant, so it could not be in thermal equilibrium with other
matter fields. The theory of cosmological phase transitions, which
was the basis for old and new inflation, did not work in such a
situation. Moreover, thermal equilibrium requires many particles
interacting with each other. This means that new inflation could
explain why our universe was so large only if it was very large
and contained many particles from the very beginning \cite{book}.

Old and new inflation represented a substantial but incomplete
modification of the big bang theory. It was still assumed that the
universe was in a state of thermal equilibrium from the very
beginning, that it was relatively homogeneous and large enough to
survive until the beginning of inflation, and that the stage of
inflation was just an intermediate stage of the evolution of the
universe. In the beginning of the 80's these assumptions seemed
most natural and practically unavoidable. On the basis of all
available observations (CMB, abundance of light elements)
everybody believed that the universe was created in a hot big
bang. That is why it was so difficult to overcome a certain
psychological barrier and abandon all of these assumptions. This
was done in 1983 with the invention of the chaotic inflation scenario
\cite{Chaot}. This scenario resolved all problems of old and new
inflation. According to this scenario, inflation may begin even if there was no thermal
equilibrium in the early universe, and  it may occur even in
the theories with simplest potentials such as $V(\phi) \sim
\phi^2$. But it is not limited to the theories with polynomial potentials: chaotic inflation occurs in {\it any} theory where the potential has a sufficiently flat region, which allows the existence of the slow-roll regime \cite{Chaot}.

\section{Chaotic Inflation}

\subsection{Basic model}

Consider  the simplest model of a scalar field $\phi$ with a mass
$m$ and with the potential energy density $V(\phi)  = {m^2\over 2}
\phi^2$. Since this function has a minimum at $\phi = 0$,  one may
expect that the scalar field $\phi$ should oscillate near this
minimum. This is indeed the case if the universe does not expand,
in which case equation of motion for the scalar field  coincides
with equation for harmonic oscillator, $\ddot\phi = -m^2\phi$.

However, because of the expansion of the universe with Hubble
constant $H = \dot a/a $, an additional  term $3H\dot\phi$ appears
in the harmonic oscillator equation:
\begin{equation}\label{1x}
 \ddot\phi + 3H\dot\phi = -m^2\phi \ .
\end{equation}
The term $3H\dot\phi$ can be interpreted as a friction term. The
Einstein equation for a homogeneous universe containing scalar
field $\phi$ looks as follows:
\begin{equation}\label{2x}
H^2 +{k\over a^2} ={1\over 6}\, \left(\dot \phi ^2+m^2 \phi^2)
\right) \ .
\end{equation}
Here $k = -1, 0, 1$ for an open, flat or closed universe
respectively. We work in units $M_p^{-2} = 8\pi G = 1$.

If   the scalar field $\phi$  initially was large,   the Hubble
parameter $H$ was large too, according to the second equation.
This means that the friction term $3H\dot\phi$ was very large, and
therefore    the scalar field was moving   very slowly, as a ball
in a viscous liquid. Therefore at this stage the energy density of
the scalar field, unlike the  density of ordinary matter, remained
almost constant, and expansion of the universe continued with a
much greater speed than in the old cosmological theory. Due to the
rapid growth of the scale of the universe and a slow motion of the
field $\phi$, soon after the beginning of this regime one has
$\ddot\phi \ll 3H\dot\phi$, $H^2 \gg {k\over a^2}$, $ \dot \phi
^2\ll m^2\phi^2$, so  the system of equations can be simplified:
\begin{equation}\label{E04}
H= {\dot a \over a}   ={ m\phi\over \sqrt 6}\ , ~~~~~~  \dot\phi =
-m\  \sqrt{2\over 3}     .
\end{equation}
The first equation shows that if the field $\phi$ changes slowly,
the size of the universe in this regime grows approximately as
$e^{Ht}$, where $H = {m\phi\over\sqrt 6}$. This is the stage of
inflation, which ends when the field $\phi$ becomes much smaller
than $M_p=1$. Solution of these equations shows that after a long
stage of inflation  the universe initially filled with the field
$\phi   \gg 1$  grows  exponentially \cite{book}, 
\begin{equation}\label{E04aa}
 a= a_0
\ e^{\phi^2/4} \   .
\end{equation}

Thus, inflation does not require initial state of thermal equilibrium, 
supercooling and tunneling from the false  vacuum. It appears in the theories that can be as simple as a theory of a harmonic oscillator \cite{Chaot}. Only when it was
realized, it became clear  that inflation is not just a trick
necessary to fix  problems of the old big bang theory, but a
generic cosmological regime.

\begin{figure}
\centering{\hskip 0.5cm
\includegraphics[height=10cm]{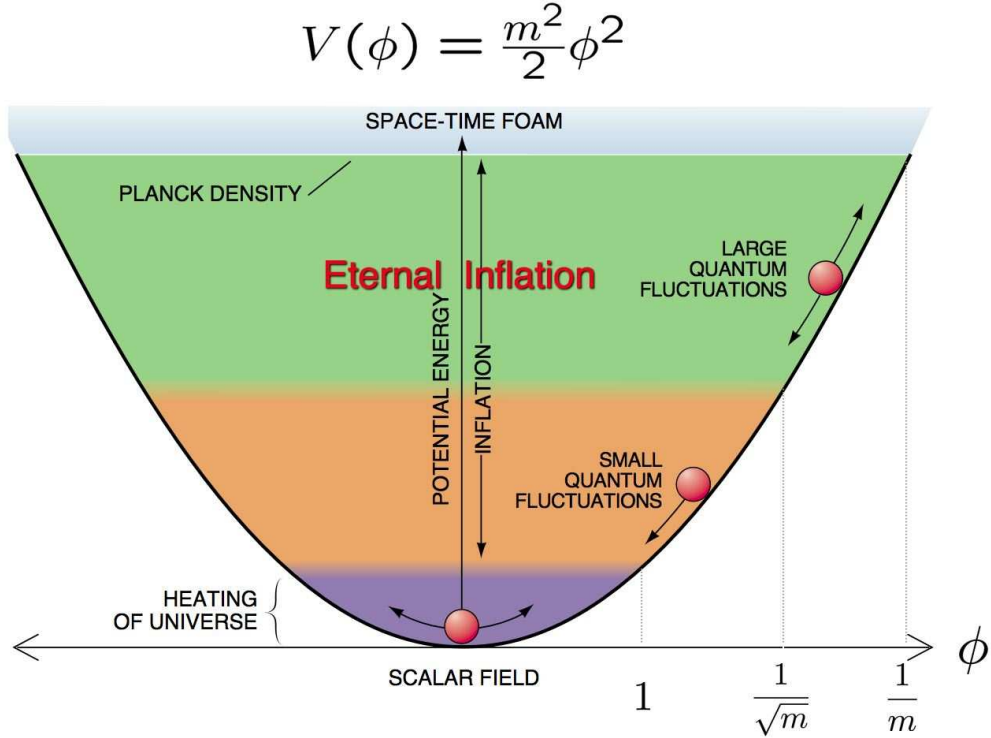}}

\

\

\caption{Motion of the scalar field in the theory with $V(\phi) =
{m^2\over 2} \phi^2$. Several different regimes are possible,  depending
on the value of the field $\phi$. If the potential energy density of the
field is greater than the Planck density $M_p^4 = 1$, $\phi \gtrsim m^{-1}$,
quantum fluctuations of space-time are so strong that one cannot describe
it in usual terms. Such a state   is called space-time foam. At a
somewhat smaller energy density  (for $m  \lesssim V(\phi) \lesssim 1$,\, $m^{-1/2} \lesssim \phi \lesssim  m^{-1}$) quantum fluctuations of space-time are small, but quantum fluctuations of
the scalar field $\phi$ may be large. Jumps of the scalar field due to
quantum fluctuations lead to a process of eternal self-reproduction of
inflationary universe which we are going to discuss later. At even
smaller values of $V(\phi)$ (for $m^2  \lesssim V(\phi) \lesssim m$,\, $1 \lesssim \phi \lesssim  m^{-1/2}$)
fluctuations of the field $\phi$ are small; it slowly moves down as a
ball in a viscous liquid. Inflation occurs for $1 \lesssim \phi \lesssim  m^{-1}$. Finally, near the minimum of $V(\phi)$ (for $\phi \lesssim 1$) the scalar
field rapidly oscillates, creates pairs of elementary particles, and the
universe becomes hot.} \label{fig:Fig1}
\end{figure}

\subsection{Initial conditions}

But what is about the initial conditions required for chaotic inflation?
Let us consider first a closed universe of initial  size $l \sim 1$
(in Planck units), which emerges  from the
space-time foam, or from singularity, or from `nothing'  in a state
with  the Planck density $\rho \sim 1$. Only starting from this moment, i.e. at $\rho
\lesssim 1$, can we describe this domain as  a {\it classical} universe.  Thus,
at this initial moment the sum of the kinetic energy density, gradient energy
density, and the potential energy density  is of the order unity:\, ${1\over
2} \dot\phi^2 + {1\over 2} (\partial_i\phi)^2 +V(\phi) \sim 1$.

We wish to emphasize, that there are no {\it a priori} constraints on
the initial value of the scalar field in this domain, except for the
constraint ${1\over 2} \dot\phi^2 + {1\over 2} (\partial_i\phi)^2 +V(\phi) \sim
1$.  Let us consider for a moment a theory with $V(\phi) = const$. This theory
is invariant under the {\it shift symmetry}  $\phi\to \phi + c$. Therefore, in such a
theory {\it all} initial values of the homogeneous component of the scalar field
$\phi$ are equally probable.  

The only constraint on the  amplitude of the field appears if the
effective potential is not constant, but grows and becomes greater
than the Planck density at $\phi > \phi_p$, where  $V(\phi_p) = 1$. This
constraint implies that $\phi \lesssim \phi_p$,
but there is no reason to expect that initially
$\phi$ must be much smaller than $\phi_p$. This suggests that the  typical initial value
 of the field $\phi$ in such a  theory is  $\phi 
\sim  \phi_p$.

Thus, we expect that typical initial conditions correspond to
${1\over 2}
\dot\phi^2 \sim {1\over 2} (\partial_i\phi)^2\sim V(\phi) = O(1)$.
If ${1\over 2} \dot\phi^2 + {1\over 2} (\partial_i\phi)^2
 \lesssim V(\phi)$
in the domain under consideration, then inflation begins,
and then within
the Planck time the terms  ${1\over 2} \dot\phi^2$ and ${1\over 2}
(\partial_i\phi)^2$ become much smaller than $V(\phi)$, which ensures
continuation of inflation.  It seems therefore that chaotic inflation
occurs under rather natural initial conditions, if it can begin at $V(\phi)
\sim 1$ \cite{Linde:1985ub,book}. 

One can get a different perspective on this issue by studying the probability of quantum creation of the universe from ``nothing.'' The basic idea is that quantum fluctuations can create a small universe from nothing if it can be done quickly, in agreement with the quantum uncertainty relation $\Delta E\cdot \Delta t \lesssim 1$. The total energy of scalar field in a closed inflationary universe is proportional to its minimal volume $H^{-3} \sim V^{-3/2}$ multiplied by the energy density $V(\phi)$:~ $E \sim V^{-1/2}$.  Therefore such a universe can appear quantum mechanically within the time $ \Delta t \gtrsim 1$ if  $V(\phi)$ is not too much smaller than the Planck density $O(1)$. 

This qualitative conclusion agrees with the result of the investigation in the context of quantum cosmology. Indeed, according to \cite{Linde:1983mx,Vilenkin:1984wp}, the probability of quantum creation of a closed universe is proportional to
\be
P \sim \exp \left(-{24\pi^{2}\over V}\right)\ ,
\ee
which means that the universe can be created if $V$ is not too much smaller than the Planck density. The Euclidean approach to the quantum creation of the universe is based on the analytical continuation of the Euclidean de Sitter solution to the real time. This continuation is possible if $\dot \phi = 0$ at the moment of quantum creation of the universe. Thus in the simplest chaotic inflation model with $V(\phi)  = {m^2\over 2}
\phi^2$ the universe  is created in a state with $V(\phi) \sim 1$, $\phi \sim m^{-1} \gg 1$ and $\dot \phi = 0$, which is a perfect initial condition for inflation in this model \cite{Linde:1983mx,book}.

One should note that there are many other attempts to evaluate the probability of initial conditions for inflation. For example, if one interprets the square of the  Hartle-Hawking wave function \cite{Hartle:1983ai} as a probability of initial condition, one obtains a paradoxical answer $P \sim \exp \left({24\pi^{2}\over V}\right)$, which could seem to imply that it is easier to create the universe with  $V\to 0$ and with an infinitely large total energy $E \sim V^{-1/2} \to \infty$. There were many attempts to improve this anti-intuitive answer, but from my perspective these attempts were misplaced: The Hartle-Hawking wave function was derived in  \cite{Hartle:1983ai} as a wave function for the {\it ground state of the universe}, and therefore it describes the most probable {\it final} state of the universe, instead of the probability of initial conditions; see a discussion of this issue in \cite{Open,book,Linde:2006nw}.

Another recent attempt to study this problem was made by Gibbons and Turok  \cite{Gibbons:2006pa}. They studied classical solutions describing a combined evolution of a scalar field and the scale factor of the universe, and imposed ``initial conditions'' not at the beginning of inflation but at its end. Since one can always reverse the direction of time in the solutions, one can always relate the conditions at the end of inflation to the conditions at its beginning. If one assumes that certain conditions at the end of inflation are equally probable, then one may conclude that the probability of initial conditions suitable for inflation must be very small \cite{Gibbons:2006pa}.  

From our perspective \cite{Kofman:2002cj,koflingibbtur}, we have here the same paradox which is encountered in the discussion of the  growth of entropy. If one starts with a well ordered system, its entropy will always grow. However, if we make a movie of this process, and play it back starting from the end of the process, then the final conditions for the original system become the initial conditions for the time-reversed system, and we will see the entropy decreasing. That is why replacing initial conditions by final conditions can be very misleading. An advantage of the inflationary regime is that it is an attractor (i.e. the most probable regime) for the family of solutions describing an expanding universe. But if one replaces initial conditions by the final conditions at the end of the process and then studies the same process back in time, the same trajectory will look like a repulsor. This is the main reason of the negative conclusion of Ref.  \cite{Gibbons:2006pa}.

The main problem with  \cite{Gibbons:2006pa} is that the methods developed there are valid for the classical evolution of the universe, but the initial conditions for the classical evolution are determined by the processes at the quantum epoch near the singularity, where the methods of  \cite{Gibbons:2006pa} are inapplicable. It is not  surprising, therefore,  that the results of \cite{Gibbons:2006pa} imply that initially $\dot\phi^{2} \gg V(\phi)$. This result contradicts the results of the Euclidean approach to  quantum creation of the universe \cite{Linde:1983mx,Vilenkin:1984wp,Hartle:1983ai} which require that initially $\dot \phi = 0$, see a discussion above.

As we will show in a separate publication \cite{koflingibbtur}, if one further develops the  methods  of \cite{Gibbons:2006pa}, but imposes the initial conditions at the beginning of inflation, rather than at its end, one finds that inflation is most probable, in agreement with the arguments given in the first part of this section. 

The discussion of initial conditions in this section was limited to the simplest versions of chaotic inflation which allow inflation at the very high energy densities, such as the models with $V\sim \phi^{n}$. We will return to the discussion of the problem of initial conditions in inflationary cosmology in Sections \ref{torus} and \ref{land}, where we will analyze it in the context of more complicated inflationary models.

\subsection{Solving the cosmological problems}

As we will see shortly, the realistic value of the mass $m$ is about $3\times 10^{-6}$, in Planck units. Therefore, according to Eq. (\ref{E04aa}), the total amount of inflation achieved starting from $V(\phi) \sim 1$ is of the order $10^{10^{10}}$. The total duration of
inflation in this model is about $10^{-30}$ seconds. When inflation
ends, the scalar field $\phi$ begins to   oscillate near the
minimum of $V(\phi)$. As any rapidly oscillating classical field,
it looses its energy by creating pairs of elementary particles.
These particles interact with each other and come to a state of
thermal equilibrium with some temperature $T_{r}$ \cite{oldtheory,KLS,tach,Desroche:2005yt,latticeold,latticeeasy,thermalization}.
From this time on, the universe can be described by the usual big
bang theory.

The main difference between inflationary theory and the old
cosmology becomes clear when one calculates the size of a typical
inflationary domain at the end of inflation. Investigation of this
question    shows that even if  the initial size of   inflationary
universe  was as small as the Planck size $l_P \sim 10^{-33}$ cm,
after $10^{-30}$ seconds of inflation   the universe acquires a
huge size of   $l \sim 10^{10^{10}}$ cm! This number is
model-dependent, but in all realistic models the  size of the
universe after inflation appears to be many orders of magnitude
greater than the size of the part of the universe which we can see
now, $l \sim 10^{28}$ cm. This immediately solves most of the
problems of the old cosmological theory \cite{Chaot,book}.

Our universe is almost exactly homogeneous on  large scale because
all inhomogeneities were exponentially stretched during inflation.
The density of  primordial monopoles  and other undesirable
``defects'' becomes exponentially diluted by inflation.   The
universe   becomes enormously large. Even if it was a closed
universe of a size
 $\sim 10^{-33}$ cm, after inflation the distance between its ``South'' and
``North'' poles becomes many orders of magnitude greater than
$10^{28}$ cm. We see only a tiny part of the huge cosmic balloon.
That is why nobody  has ever seen how parallel lines cross. That
is why the universe looks so flat.

If our universe initially consisted of many domains with
chaotically distributed scalar field  $\phi$ (or if one considers
different universes with different values of the field), then
domains in which the scalar field was too small never inflated.
The main contribution to the total volume of the universe will be
given by those domains which originally contained large scalar
field $\phi$. Inflation of such domains creates huge homogeneous
islands out of initial chaos. (That is why I called this scenario
``chaotic inflation.'') Each  homogeneous domain in this scenario
is much greater than the size of the observable part of the
universe.

\subsection{Chaotic inflation versus new inflation}

The first models of chaotic inflation were based on the theories
with polynomial potentials, such as $V(\phi) = \pm {m^2\over 2}
\phi^2 +{\lambda\over 4} \phi^4$. But, as was emphasized in  \cite{Chaot}, the main idea of this scenario is quite generic. One should consider any particular
potential $V(\phi)$, polynomial or not, with or without
spontaneous symmetry breaking, and study all possible initial
conditions without assuming that the universe was in a state of
thermal equilibrium, and that the field $\phi$ was in the minimum
of its effective potential from the very beginning.

This scenario strongly deviated from the standard lore of the hot
big bang theory and was psychologically difficult to accept.
Therefore during the first few years after invention of chaotic
inflation many authors claimed that the idea of chaotic initial
conditions is unnatural, and made attempts to realize the new
inflation scenario based on the theory of high-temperature phase
transitions, despite numerous problems associated with it. Some
authors believed that the theory must satisfy so-called `thermal
constraints' which were necessary to ensure that the minimum of
the effective potential at large $T$ should be at $\phi=0$
\cite{OvrStein}, even though the scalar field in the models  they
considered was not in a state of thermal equilibrium with other
particles. 

The issue of thermal initial conditions played the central role in the  long debate  about new inflation versus chaotic inflation in the 80's. This debate continued for many years, and a significant part of my book \cite{book} was dedicated to it. By now  the debate is over: no realistic versions of new inflation based on the theory of thermal phase transitions and supercooling have been proposed so far. Gradually it became clear that the idea of chaotic initial conditions is most general, and
it is much easier to construct a consistent cosmological theory without making unnecessary assumptions about thermal equilibrium and high temperature phase transitions in the early universe. 

As a result, the corresponding terminology changed.  Chaotic inflation, as defined in  \cite{Chaot}, occurs in {\it all} models with sufficiently flat potentials, including the potentials with a flat maximum, originally used in new inflation \cite{Linde:cd}.  Now the versions of  inflationary scenario with such potentials for simplicity are often called `new inflation', even though inflation begins there not as in the original new inflation scenario, but as in the chaotic inflation
scenario. To avoid this terminological misunderstanding, some authors call the version of chaotic inflation scenario, where inflation occurs near the top of the scalar potential, a  `hilltop inflation' \cite{Boubekeur:2005zm}.

\section{Hybrid inflation}

The simplest models of inflation involve just one scalar field. However, in supergravity and string theory there are many different scalar fields, so it does make sense to study models with several different scalar fields, especially if they have some qualitatively new properties. Here we will consider one of these models, hybrid inflation \cite{Hybrid}.
 
 The simplest version of hybrid inflation describes the theory of two scalar fields with
the effective potential
\begin{equation}\label{hybrid}
V(\sigma,\phi) =  {1\over 4\lambda}(M^2-\lambda\sigma^2)^2 +
{m^2\over 2}\phi^2 + {g^2\over 2}\phi^2\sigma^2\ .
\end{equation}
The effective mass squared of the field $\sigma$ is equal to $-M^2
+ g^2\phi^2$.  Therefore for $\phi > \phi_c = M/g$ the only
minimum of the effective potential $V(\sigma,\phi)$ is at $\sigma
= 0$. The curvature of the effective potential in the
$\sigma$-direction is much greater than in the $\phi$-direction.
Thus  at the first stages of expansion of the universe the field
$\sigma$ rolled down to $\sigma = 0$, whereas the field $\phi$
could remain large for a much longer time.

At the moment when the inflaton field $\phi$ becomes smaller than
$\phi_c = M/g$,  the phase transition with the symmetry breaking
occurs. The fields rapidly fall to the absolute minimum of the potential at $\phi = 0, \sigma^{2 } = M^{2}/\lambda$. If $m^2 \phi_c^2 = m^2M^2/g^2 \ll M^4/\lambda$, the Hubble
constant at the time of the phase transition is given by $H^2 =
{M^4 \over 12 \lambda}$ (in units $M_p = 1$). If  
$M^2 \gg {\lambda m^2\over g^2}$ and  $m^2 \ll H^2$, then the
universe at $\phi > \phi_c$ undergoes a stage of inflation, which
abruptly ends at $\phi = \phi_c$.

Note that hybrid inflation is also a version of the chaotic inflation scenario: I am unaware of any way to realize this model in the context of the theory of high temperature phase transitions.  The main difference between this scenario and the simplest versions of the one-field chaotic inflation is in the way inflation ends. In the theory with a single field, inflation ends when the potential of this field becomes steep. In hybrid inflation, the structure of the universe depends on the way one of the fields moves, but inflation ends when the potential of the second field becomes steep. This fact allows much greater flexibility of construction of inflationary models. Several extensions of this scenario became quite popular in the context of supergravity and string cosmology, which we will discuss later.

\section{Quantum fluctuations and density perturbations
\label{Perturb}}

The average amplitude of inflationary perturbations generated during a
typical time interval $H^{-1}$ is given by
\cite{Vilenkin:wt,Linde:uu}
\begin{equation}\label{E23}
|\delta\phi(x)| \approx \frac{H}{2\pi}\ .
\end{equation}

These fluctuations lead to density perturbations that later
produce galaxies. The theory of this effect  is very complicated
\cite{Mukh,Hawk}, and it was fully understood only in the second
part of the 80's \cite{Mukh2}. The main idea can be described as
follows:

Fluctuations of the field $\phi$ lead to a local delay of the time
of the end of inflation,  $\delta t = {\delta\phi\over \dot\phi}
\sim {H\over 2\pi \dot \phi}$. Once the usual post-inflationary
stage begins, the density of the universe starts to decrease as
$\rho = 3 H^2$, where $H \sim t^{-1}$. Therefore a local delay of
expansion leads to a local density increase $\delta_H$ such that
$\delta_H \sim \delta\rho/\rho \sim  {\delta t/t}$. Combining
these estimates together yields the famous result
\cite{Mukh,Hawk,Mukh2}
\begin{equation}\label{E24}
\delta_H \sim \frac{\delta\rho}{\rho} \sim {H^2\over 2\pi\dot\phi}
\ .
\end{equation}
The field $\phi$ during inflation
changes very slowly, so the quantity ${H^2\over 2\pi\dot\phi}$
remains almost constant over exponentially large range of
wavelengths. This means that the spectrum of perturbations of
metric is flat.

A detailed calculation in our simplest chaotic inflation model gives the following result for the amplitude of perturbations:
\begin{equation}\label{E26}
\delta_H \sim   {m \phi^2\over 5\pi \sqrt 6} \ .
\end{equation}
The perturbations on scale of the horizon were produced at
$\phi_H\sim 15$ \cite{book}. This, together  with COBE
normalization $\delta_H \sim 2 \times 10^{-5}$  gives $m \sim
3\times 10^{-6}$, in Planck units, which is approximately
equivalent to $7 \times 10^{12}$ GeV. An exact value of $m$ depends on
$\phi_H$, which in its turn depends slightly on the subsequent
thermal history of the universe.

When the fluctuations of the scalar field $\phi$ are first produced (frozen), their wavelength is given by $H(\phi)^{-1}$. At the end of inflation, the wavelength grows by the factor of $e^{\phi^2/4}$, see Eq. (\ref{E04aa}). In other words,  the logarithm of the wavelength $l$ of the perturbations of the metric is proportional to the value of $\phi^2$ at the moment when these perturbations were produced. As a result, according to  (\ref{E26}), the amplitude of perturbations of the metric depends on the wavelength  logarithmically: $\delta_H \sim   {m\, \ln l} $. A similar logarithmic dependence (with different powers of the logarithm) appears in other versions of chaotic inflation with $V \sim \phi^{n}$ and in the simplest versions of new inflation.

At the first glance, this logarithmic deviation from scale invariance could seem inconsequential, but in a certain sense it is similar to the famous logarithmic dependence of the coupling constants in QCD, where it leads to asymptotic freedom at high energies, instead of simple scaling invariance \cite{Gross:1973id,Politzer:1973fx}. In QCD, the slow growth of the coupling constants at small momenta/large distances is responsible for nonperturbative effects resulting in quark confinement. In  inflationary theory, the  slow growth of the amplitude of perturbations of the metric at large distances is equally important. It leads to the existence of the regime of eternal inflation and to the fractal structure of the universe on  super-large scales, see Section \ref{eternalinfl}.

Since the
observations provide us with  information about a rather limited
range of $l$, it is often possible to parametrize the scale dependence
of density perturbations by a simple power law, $\delta_H \sim
l^{(1-n_{s})/2}$. An exactly flat spectrum, called Harrison-Zeldovich spectrum, would correspond to $n_{s} = 1$.

The amplitude of scalar perturbations of the metric can be characterized either by $\delta_{H}$, or by a closely related quantity $\Delta_{\cal R}$ \cite{LL}. Similarly, the amplitude of tensor perturbations is given by $\Delta_h$.  Following 
\cite{LL,WMAP}, one can represent these quantities as 
\begin{eqnarray}
 \label{eq:P_R} \Delta^2_{\cal R}(k)&=& \Delta^2_{\cal R}(k_0)
  \left(\frac{k}{k_0}\right)^{n_s-1}, \\ \label{eq:P_h} \Delta^2_h(k)&=& \Delta^2_h(k_0)
  \left(\frac{k}{k_0}\right)^{n_t},
\end{eqnarray}
where $\Delta^2(k_0)$ is a normalization constant, and $k_0$ is a normalization point. Here we ignored running of the indexes $n_{s}$ and $n_{t}$ since there is no observational evidence that it is significant.

One can also introduce the tensor/scalar ratio
$r$, the relative amplitude  of the tensor to  scalar modes,
\begin{equation}
 \label{eq:rdef} r \equiv \frac{\Delta^2_h(k_0)}{\Delta^2_{\cal R}(k_0)}.
\end{equation}

There
are three slow-roll
parameters \cite{LL}
\begin{eqnarray}
 \label{eq:eps} \epsilon =
  \frac{1}{2}\left(\frac{V'}{V}\right)^2, ~~
  \eta  =  \frac{V''}{V}, ~~
  \label{eq:xi} \xi  =
\frac{V'V'''}{V^2},
\end{eqnarray}
where prime denotes derivatives with respect to the field
$\phi$. All parameters must be
smaller than one for the slow-roll approximation to be valid. 

 A standard slow roll analysis gives observable
quantities in terms of the slow roll parameters to first order as
\begin{eqnarray}
 \label{eq:A} &&\Delta^2_{\cal R}   =  \frac{V}{24\pi^2\epsilon} =  \frac{V^{3}}{12\pi^2(V')^{2}} \ ,\\
    \label{eq:n_s} &&n_s-1 =
  -6\epsilon + 2\eta = 1 
  -{3}\left(\frac{V'}{V}\right)^2 + 2\frac{V''}{V} \ , \\
  \label{eq:r} &&r   =  16 \epsilon, \\
  \label{eq:n_t} &&n_t   =  -2\epsilon = -\frac{r}{8} \ .
\end{eqnarray}
 The equation
$n_t=-r/8$ is known as the consistency relation for single-field
inflation models; it becomes an inequality for multi-field
inflation models. If $V$ during inflation is sufficiently large, as in the simplest models of chaotic inflation, one may have a chance to find the tensor contribution to the CMB anisotropy. The possibility to determine $n_{t}$ is less certain. 
The most important information which can be obtained now from the cosmological observations at present is related to Eqs. (\ref{eq:A}) and (\ref{eq:n_s}).

Following notational conventions in \cite{WMAP}, we use
$A(k_0)$ for the scalar power spectrum amplitude, where $A(k_0)$ and
$\Delta^2_{\cal R}(k_0)$ are related through
\begin{eqnarray}
 \label{eq:Adef} \Delta^2_{\cal R}(k_0) \simeq 3\times 10^{-9}  A(k_0).
\end{eqnarray}
The parameter $A$ is often normalized at $k_{0} \sim 0.05$/Mpc; its observational value is about 0.8 \cite{WMAP,Martin:2006rs,Tegmark,Kuo:2006ya}. This leads to the observational constraint on $V(\phi)$ and on $r$ following from the normalization of the spectrum of the large-scale density perturbations:
\begin{eqnarray}
 \label{eq:V} {V^{{3/2}}\over V'} \simeq 5\times 10^{-4} \ .
\end{eqnarray}
Here $V(\phi)$ should be evaluated for the value of the field $\phi$ which is determined by the condition that the perturbations produced at the moment when the field was equal $\phi$  evolve into the present time perturbations with momentum $k_{0} \sim 0.05$/Mpc. In the first approximation, one can find the corresponding moment by assuming that it happened 60 e-foldings before the end of inflation. The number of e-foldings can be calculated in the slow roll approximation using the relation
\begin{eqnarray}
 \label{eq:N} {N} \simeq \int_{\phi_{\rm end}}^{\phi}{V\over V'} d\phi \ .
\end{eqnarray}
Equation (\ref{eq:V})  leads to the relation between $r$, $V$ and $H$, in Planck units:
\begin{eqnarray}
 \label{eq:rvh} {r} \approx 3 \times  10^{7}~V  \approx 10^{8}~ H^{2} \ .
\end{eqnarray}
Recent observational data show that $r \lesssim 0.3$ and
\begin{eqnarray}
 \label{eq:ns} n_s = 0.95 \pm  0.016  
\end{eqnarray}
for $r \ll 0.1$  \cite{Tegmark}.  These relations are very useful for comparing inflationary models with observations.
In particular, the simplest versions of chaotic and new inflation predict $n_{s} < 1$, whereas in hybrid inflation one may have either $n_{s} < 1$ or $n_{s}>  1$, depending on the model. A more accurate representation of observational constraints can be found in Section \ref{observ}.

Until now we discussed the standard mechanism of generation of perturbations of the metric. However, if the model is sufficiently complicated, other mechanisms become possible. For example, one may consider a theory of two  scalar fields, $\phi$ and $\sigma$, and assume that   inflation was driven by the field $\phi$, and the field $\sigma$ was very light during inflation and did not contribute much to the total energy density. Therefore its quantum fluctuations also did not contribute much to the amplitude of perturbations of the metric during inflation  (isocurvature perturbations).

After inflation the field $\phi$ decays. If the products of its decay rapidly loose energy, the field $\sigma$ may dominate the energy density of the universe and its perturbations suddenly become important. If, in its turn, the field $\sigma$ decays, its perturbations under certain conditions can be  converted into the usual adiabatic perturbations of the metric. If this conversion is incomplete, one obtains a theory  at odds with recent observational data \cite{Ax,Mollerach:1989hu}. On the other hand, if the conversion is complete, one obtains a novel mechanism of generation of adiabatic density perturbations, which is called the curvaton mechanism \cite{LM,Enqvist:2001zp,LW,Moroi:2001ct}. A closely related
but different mechanism was also proposed in \cite{mod1}. For a recent review, see \cite{Wands:2007bd}.

These mechanisms are much more complicated than the original one, but one should keep them in mind since they sometimes work in the situations where the standard one does not. Therefore they can give us an additional freedom in finding realistic models of inflationary cosmology.

\section{Creation of matter after inflation: reheating and preheating}

The theory of reheating of the universe after inflation is the most
important application of the quantum theory of particle creation, since
almost all matter constituting the universe  was created during this
process.

At the stage of inflation all energy is concentrated in a classical
slowly moving inflaton field $\phi$. Soon after the end of inflation this
field begins to oscillate near the minimum of its effective potential.
Eventually it produces many elementary particles, they interact  with
each other and come to a state of thermal equilibrium with some
temperature $T_r$.

Early discussions of reheating of the universe after inflation  \cite{oldtheory} were based on the idea that the homogeneous inflaton field can be represented as a collection of the particles of the field $\phi$. Each of these particles decayed independently. This process can be studied by the usual perturbative approach to particle decay. Typically, it takes thousands of oscillations of the inflaton field until it decays into usual elementary particles by this mechanism.  More recently, however, it was discovered that coherent field effects such as parametric resonance can lead to the decay of the homogeneous field much faster than would have been predicted by perturbative methods, within few dozen oscillations \cite{KLS}. These coherent effects produce high energy, nonthermal fluctuations that could have
significance for understanding developments in the early universe, such as baryogenesis.  This early stage of rapid nonperturbative decay  was called `preheating.' 
In \cite{tach} it was found that another effect known as tachyonic preheating can lead to even faster decay than parametric
resonance. This effect occurs whenever the homogeneous field rolls down a tachyonic ($V''<0$) region of its
potential. When that occurs, a tachyonic, or spinodal instability leads to exponentially rapid growth of all long wavelength modes with 
$k^2<|V''|$. This growth can often drain all of the energy from the homogeneous field within a single oscillation.

We are now in a position to classify the dominant mechanisms by which the homogeneous inflaton field decays in different classes of
inflationary models. Even though all of these models, strictly speaking,  belong to the general class of chaotic inflation (none of them is based on the theory of thermal initial conditions), one can  break them into three classes: small field, or new inflation models \cite{New}, large field, or chaotic inflation models of the type of the model $m^2\phi^2/2$ \cite{Chaot}, and multi-field, or hybrid models \cite{Hybrid}. This classification is incomplete, but still rather helpful.

In the simplest versions of chaotic inflation, the stage of preheating is generally dominated by
parametric resonance, although there are parameter ranges where this
can not occur \cite{KLS}.    In \cite{tach} it was shown that tachyonic preheating
dominates the preheating phase in hybrid models of inflation. New inflation in this respect occupies an intermediate position between chaotic inflation and hybrid inflation:  If spontaneous symmetry breaking in this scenario is very large, reheating occurs due to parametric resonance and perturbative decay. However, for the models with spontaneous symmetry breaking at or below the GUT scale, $\phi \ll 10^{{-2}} M_p$, preheating occurs due to a combination of tachyonic preheating and parametric resonance. The resulting effect is very strong, so that the homogeneous mode of the inflaton field typically decays within few oscillations \cite{Desroche:2005yt}. 

A detailed investigation of preheating usually requires lattice simulations, which can be achieved following \cite{latticeold,latticeeasy}.  Note that preheating is not the last stage of reheating; it is followed by a period of turbulence \cite{thermalization}, by a much slower perturbative decay described by the methods developed in \cite{oldtheory}, and by eventual thermalization.

\section{Eternal inflation}\label{eternalinfl}

A significant step in the development of inflationary theory was
the discovery of the process of self-reproduction of inflationary
universe. This process was known to exist in old inflationary
theory \cite{Guth} and in the new one \cite{StLin,linde1982,Vilenkin:xq}, but its
significance was fully realized only after the discovery of the
regime of eternal inflation in chaotic inflation scenario \cite{Eternal,LLM}. It appears that in many inflationary models large quantum fluctuations produced during inflation
may significantly increase the value of the energy density in some
parts of the universe. These regions expand at a greater rate than
their parent domains, and quantum fluctuations inside them lead to
production of new inflationary domains which expand even faster.
This  leads to an eternal process of self-reproduction of the
universe. Most importantly, this process may divide the universe into exponentially many exponentially large parts with different laws of low-energy physics operating in each of them. The universe becomes an {\it inflationary multiverse} \cite{Eternal,LLM} (see also \cite{linde1982,nuff}).

To understand the mechanism of self-reproduction one should
remember that the processes separated by distances $l$ greater
than $H^{-1}$ proceed independently of one another. This is so
because during exponential expansion the distance between any two
objects separated by more than $H^{-1}$ is growing with a speed
exceeding the speed of light. As a result, an observer in the
inflationary universe can see only the processes occurring inside
the horizon of the radius  $H^{-1}$. An important consequence of
this general result is that the process of inflation in any
spatial domain of radius $H^{-1}$ occurs independently of any
events outside it. In this sense any inflationary domain of
initial radius exceeding $H^{-1}$ can be considered as a separate
mini-universe.

To investigate the behavior of such a mini-universe, with an
account taken of quantum fluctuations, let us consider an
inflationary domain of initial radius $H^{-1}$ containing
sufficiently homogeneous field with initial value $\phi \gg M_p$.
Equation (\ref{E04}) implies that during a typical time interval
$\Delta t=H^{-1}$ the field inside this domain will be reduced by
$\Delta\phi = \frac{2}{\phi}$. By comparison this expression with
$|\delta\phi(x)| \approx \frac{H}{2\pi} =  {m\phi\over 2\pi\sqrt
6}$ one can easily see that if $\phi$ is much less than $\phi^*
\sim {5\over  \sqrt{ m}} $,
 then the decrease of the field $\phi$
due to its classical motion is much greater than the average
amplitude of the quantum fluctuations $\delta\phi$ generated
during the same time. But for   $\phi \gg \phi^*$ one has
$\delta\phi (x) \gg \Delta\phi$. Because the typical wavelength of
the fluctuations $\delta\phi (x)$ generated during the time is
$H^{-1}$, the whole domain after $\Delta t = H^{-1}$ effectively
becomes divided into $e^3 \sim 20$ separate domains
(mini-universes) of radius $H^{-1}$, each containing almost
homogeneous field $\phi - \Delta\phi+\delta\phi$.   In almost a
half of these domains the field $\phi$ grows by
$|\delta\phi(x)|-\Delta\phi \approx |\delta\phi (x)| = H/2\pi$,
rather than decreases. This means that the total volume of the
universe containing {\it growing} field $\phi$ increases 10 times.
During the next time interval $\Delta t = H^{-1}$ this process
repeats. Thus, after the two time  intervals $H^{-1}$ the total
volume of the universe containing the growing scalar field
increases 100 times, etc. The universe enters eternal process of
self-reproduction.

The existence of this process implies that the  universe will never disappear as a whole. Some of its parts may collapse, the life in our part of the universe may perish, but there always will be some other parts of the universe where life will appear again and again, in all of its possible forms.

One should be careful, however, with the interpretation of these results. There is still an ongoing debate of whether eternal inflation is eternal only in the future or also in the past. In order to understand what is going on, let us consider any particular time-like geodesic line at the stage of inflation. One can show that for any given observer following this geodesic, the duration $t_{i}$ of the stage of inflation on this geodesic will be finite. One the other hand, eternal inflation implies that if one takes all such geodesics and calculate the time $t_{i}$ for each of them, then there will be no upper bound for $t_{i}$, i.e. for each time $T$ there will be such geodesic which experience inflation for the time $t_{i} >T$.  Even though the relative number of long geodesics can be very small, exponential expansion of space surrounding them will lead to an eternal exponential growth of the total volume of inflationary parts of the universe.

Similarly, if one concentrates on any particular geodesic in the past time direction, one can prove that it has finite length \cite{Borde:2001nh}, i.e. inflation in  any particular point of the universe should have a beginning at some time $\tau_{i}$. However, there is no reason to expect that there is an upper bound for all $\tau_{i}$ on all geodesics. If this upper bound does not exist, then eternal inflation is eternal not only in the future but also in the past.

In other words, there was a  beginning for each part of the universe, and there will be an end for inflation at any particular point. But there will be no end for the evolution of the universe {\it as a whole} in the eternal inflation scenario, and at present we do not have any reason to believe that there was a single beginning of the evolution of the whole universe at some moment $t = 0$, which was traditionally associated with the Big Bang.

To illustrate the process of eternal inflation, we present here the results of computer simulations of evolution of a system of two scalar fields during inflation. The field $\phi$ is the inflaton field driving inflation; it is shown by the height of the distribution of the field $\phi(x,y)$ in a two-dimensional slice of the universe. The second field, $\Phi$, determines the type of spontaneous symmetry breaking which may occur in the theory. We paint the surface  red, green or blue corresponding to three different minima of the potential of the field $\Phi$. Different colors correspond to different types of spontaneous symmetry breaking, and therefore to different sets of laws of low-energy physics in different exponentially large parts of the universe.

\begin{figure}

\centering\leavevmode\epsfysize=13cm \epsfbox{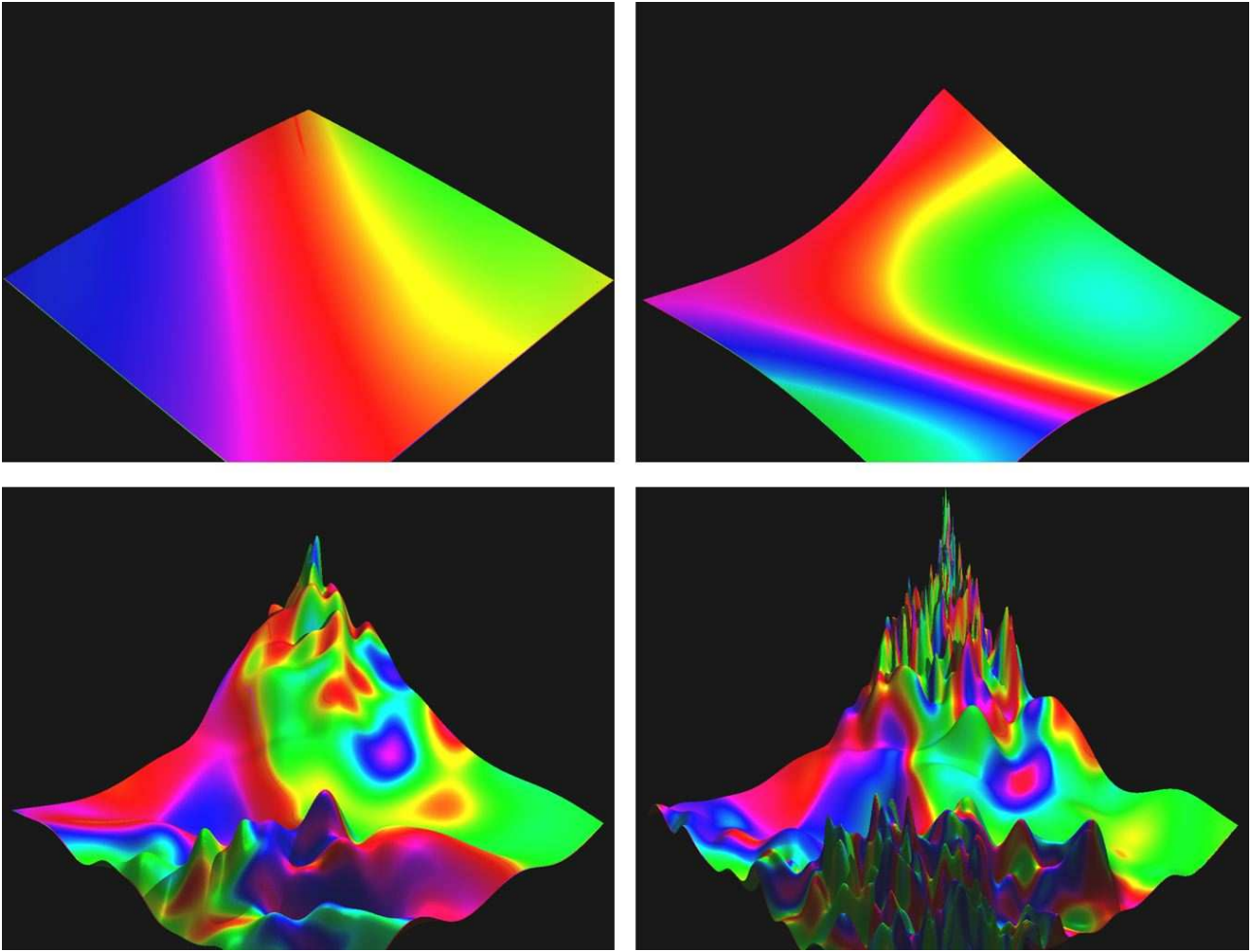}

\

\caption{Evolution of scalar fields  $\phi$ and $\Phi$ during the process of self-reproduction of the universe.   The height of the distribution shows the value of the field $\phi$ which drives inflation. The surface is painted red, green or blue corresponding to three different minima of the potential of the field $\Phi$.   Laws of low-energy physics are different in the regions of different color. The peaks of the ``mountains'' correspond  to places where quantum fluctuations bring the scalar fields back to the Planck density. Each of such places in a certain sense can be considered as a beginning of a new Big Bang. At the end of inflation, each such part becomes exponentially large. The universe becomes a multiverse,   a huge eternally growing fractal  consisting of different exponentially large  locally homogeneous parts with different laws of low-energy physics operating in each of them.}
\label{fig:Fig0}
\end{figure}

In the beginning of the process the whole inflationary domain is red, and the distribution of both fields is very homogeneous. Then the domain became exponentially large (but it has the same size in comoving coordinates, as shown in Fig. \ref{fig:Fig0}).  Each peak of the mountains corresponds to nearly Planckian density and can be interpreted as a beginning of a new ``Big Bang.'' The laws of physics are rapidly changing there, as indicated by changing colors, but they become fixed in the parts of the universe where the field $\phi$ becomes small. These parts correspond to valleys in Fig. \ref{fig:Fig0}. Thus quantum fluctuations of the scalar fields divide the universe into exponentially large domains with different laws of low-energy physics, and with different values of energy density. This makes our universe look as a multiverse, a collection of different exponentially large inflationary universes \cite{Eternal,LLM}.

Eternal inflation scenario was extensively studied during the last 20 years. I should mention, in particular, the discovery of the topological eternal inflation \cite{TopInf}  and the calculation of the fractal dimension of the universe \cite{Aryal:1987vn,LLM}. The most interesting recent developments of the theory of eternal inflation are related to the theory of inflationary multiverse and string theory landscape. We will discuss these subjects   in   Section \ref{land}.

\section{Inflation and observations}\label{observ}

Inflation is not just an interesting theory that can resolve many
difficult problems of the standard  Big Bang cosmology. This
theory made several   predictions which can be tested by
cosmological observations. Here are the most important
predictions:

1) The universe must be flat. In most models $\Omega_{total} = 1
\pm 10^{-4}$.

2) Perturbations of the metric produced during inflation are
adiabatic.

3) These perturbations are gaussian.

4)  Inflationary perturbations generated during a  slow-roll regime with $\epsilon,\eta \ll 1$   have a nearly flat spectrum with $n_{s}$ close to 1.

5) On the other hand, the spectrum of inflationary perturbations usually is slightly non-flat. It is possible to construct models with  $n_{s}$ extremely close to 1, or even exactly equal to 1. However, in general, the small deviation of the spectrum from the exact flatness is one of the distinguishing features of inflation. It is as significant for inflationary theory as the asymptotic freedom for the theory of strong interactions.

6) perturbations of the metric could be scalar, vector or tensor.
Inflation mostly produces scalar  perturbations, but it also
produces tensor perturbations with nearly flat spectrum, and it
does {\it not} produce vector perturbations. There are certain
relations between the properties of  scalar and tensor
perturbations produced by inflation.

7) Inflationary perturbations produce specific peaks in the
spectrum of CMB radiation. (For a simple pedagogical
interpretation of this effect see  e.g. \cite{Dodelson:2003ip}; a
detailed theoretical description can be found in
\cite{Mukhanov:2003xr}.)

It is possible to violate each of these predictions if one makes
inflationary theory sufficiently complicated. For example, it is possible
to produce vector perturbations of the metric in the models where
cosmic strings are produced at the end of inflation, which is the
case in some versions of hybrid inflation. It is possible to have
an open or closed inflationary universe, or even a small periodic
inflationary universe, it is possible to have models with
nongaussian isocurvature fluctuations with a non-flat spectrum.
However, it is difficult to do so, and most of the
inflationary models obey the simple rules given above.

It is not easy to test all of these predictions. The major
breakthrough in this direction was achieved  due to the recent
measurements of the CMB anisotropy. The latest results based on
the WMAP experiment, in combination with the Sloan Digital Sky
Survey, are consistent with predictions of the simplest
inflationary models with adiabatic gaussian perturbations, with
$\Omega = 1.003 \pm 0.01$, and $n_{s} = 0.95 \pm
0.016$~\cite{Tegmark}.

\begin{figure}

\centering\leavevmode\epsfysize=8cm \epsfbox{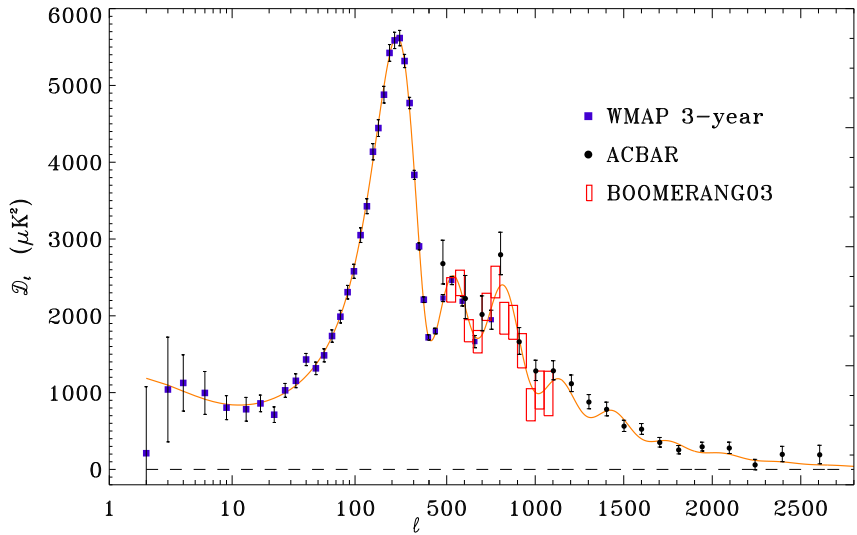}

\caption{CMB data (WMAP3, BOOMERANG03, ACBAR) versus the predictions of one of the simplest inflationary models with $\Omega = 1$ (red line), according to \cite{Kuo:2006ya}.}
\label{cmb}
\end{figure}

There are still some question marks to be examined, such as an
unexpectedly small anisotropy of CMB at  large angles \cite{WMAP,Efstathiou:2003wr} and possible correlations between low multipoles; for a recent discussion see e.g. \cite{Land:2006bn,Rakic:2007ve} and references therein.

The observational status and interpretation of these effects is still uncertain, but if one takes these effects seriously, one may try to look for some theoretical explanations. For example, there are several ways to suppress the large angle anisotropy, see e.g. \cite{Contaldi}. The situation with correlations between low multipoles requires more work. In particular, it would be interesting to study effects related to relatively light domain walls \cite{Stebbins:1988bs,Turner:1990uw,Battye:2006mb}. Another possibility is to analyze the possible effects on the CMB anisotropy which can be produced by the cosmic web structure of the perturbations in the curvaton scenario \cite{LM}. Some other possibilities are mentioned in \cite{Rakic:2007ve}. One way or another,  it is quite significant that all proposed explanations of these anomalies are based on inflationary cosmology.

One of the interesting issues to be probed by the future observations is the possible existence of gravitational waves produced during inflation.  The present upper bound on the tensor to scalar ratio $r$  is not very strict, $r \lesssim 0.3$. However, the new observations may either find the tensor modes or  push the bound on $r$ much further, towards $r\lesssim 10^{-2}$ or even $r\lesssim 10^{-3}$. 

In the simplest monomial versions of chaotic inflation with $V \sim \phi^{n}$ one find the following (approximate) result: $r = 4n/N$. Here N is the number of e-folds of inflation corresponding to the wavelength equal to the present size of the observable part of our universe; typically $N$ can be in the range of 50 to 60; its value depends on the mechanism of reheating. For the simplest model with $n = 2$ and $N \sim 60$ one has $r \sim 0.13 - 0.14$. On the other hand, for most of the other models, including the original version of new inflation, hybrid inflation, and many versions of string theory inflation, $r$ is extremely small, which makes observation of gravitational waves in such models  very difficult. 

One may wonder whether there are any sufficiently simple and natural models with intermediate values of $r$? This is an important questions for those who are planning a new generation of the CMB experiments. The answer to this question is positive: In the versions of chaotic inflation with potentials like $\pm m^{2}\phi^{2} +\lambda\phi^{4}$, as well as in the natural inflation scenario, one can easily  obtain any value of $r$ from $0.3$ to $10^{-2}$. I will illustrate it with two figures. The first one shows the graph of possible values of $n_{s}$ and $r$ in the standard symmetry breaking model  with the potential 
\begin{equation}
V = - m^{2}\phi^{2}/2 +\lambda\phi^{4}/4 + m^{4}/4\lambda =  {\lambda\over 4}(\phi^{2}-v^{2})^{2} \ ,
\end{equation}
where $v = m/\sqrt\lambda$ is the amplitude of spontaneous symmetry breaking.  
\begin{figure}[h!]
\centering\leavevmode\epsfysize=5cm \epsfbox{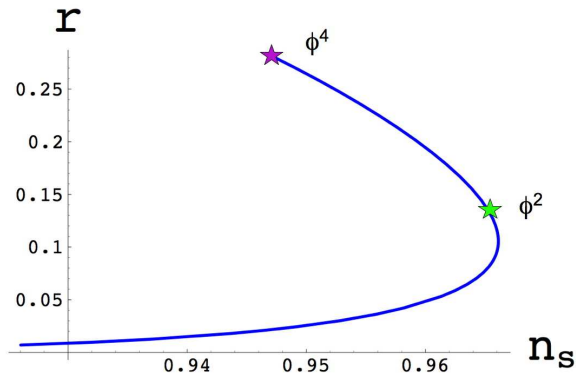}  
\caption{Possible values of $r$ and $n_{s}$ in the theory ${\lambda\over 4}(\phi^{2}-v^{2})^{2}$ for different initial conditions and different $v$, for $N = 60$.  In the small $v$ limit, the model has the same predictions as the theory $\lambda\phi^{4}/4$. In the large $v$ limit it has the same predictions as the theory $m^{2}\phi^{2}$. The upper branch, above the first star from below (marked as $\phi^{2}$), corresponds to inflation which occurs while the field rolls down from  large $\phi$; the lower branch corresponds to the motion from $\phi = 0$. }
\label{cmbquadr}
\end{figure}

If $v$ is very large, $v \gtrsim 10^{2}$, inflation occurs near the minimum of the potential, and all properties of inflation are the same as in the simplest chaotic inflation model with  quadratic potential $m^{2}\phi^{2}$. If $v \ll 10$, inflation occurs as in the theory $\lambda\phi^{4}/4$, which leads to $r \sim 0.28$. If $v$ takes some intermediate values, such as $v = O(10)$, then two different inflationary regimes are possible in this model: at large $\phi$ and at small $\phi$. In the first case $r$ interpolates between its value in the theory $\lambda\phi^{4}/4$ and the theory  $m^{2}\phi^{2}$ (i.e. between $0.28$ and $0.14$). In the second case, $r$ can take any value from $0.14$ to $10^{-2}$, see Fig. \ref{cmbquadr} \cite{deVega:2006un,Kallosh:2007wm}.

\begin{figure}[h!]

\centering\leavevmode\epsfysize=7.5cm \epsfbox{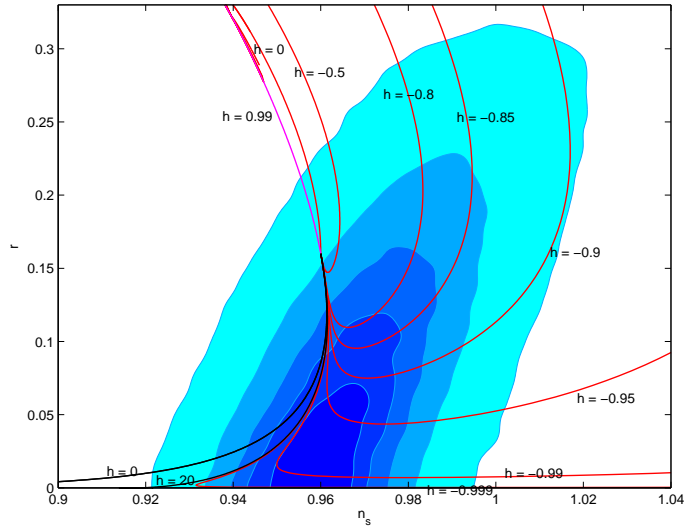}  
 
\caption{Possible values of $r$ and $n_{s}$ for chaotic inflation with a potential including terms $\phi^{2}$, $\phi^{3}$ and $\phi^{4}$ for N = 50, according to  \cite{Destri:2007pv}. The color-filled areas correspond to 12\%, 27\%, 45\%, 68\% and 95\% confidence levels according to  the WMAP3 and SDSS data. }
\label{cmbquadr2}
\end{figure}

If one considers chaotic inflation with the potential including terms $\phi^{2}$,  $\phi^{3}$ and  $\phi^{4}$, one can considerably alter the properties of inflationary perturbations \cite{Hodges:1989dw}. Depending on the values of parameters,  initial conditions and the required number of e-foldings $N$, this relatively simple class of models covers almost all parts of the area in the $(r,n_{s})$ plane allowed by the latest observational data \cite{Destri:2007pv}, see Fig. \ref{cmbquadr2}.  
 
Note that for all versions of the model shown in Figs.  \ref{cmbquadr},  \ref{cmbquadr2} the range of the cosmological evolution of the fields is $\Delta \phi > 1$, so formally these models can be called the large field models. And yet they have dramatically different properties, which do not fit into the often used scheme dividing all models into small field models, large field models and hybrid inflation models.

\section{Alternatives to inflation?}\label{alt}

Inflationary scenario is very versatile, and now, after 25 years
of persistent attempts of many physicists to propose an
alternative to inflation, we still do not know any other  way to
construct a consistent cosmological theory. Indeed, in order to
compete with inflation a new theory should make similar
predictions and should offer an alternative solution to many
difficult cosmological problems. Let us look at these problems
before starting a discussion.

1) Homogeneity problem. Before even starting investigation of
density perturbations and structure  formation, one should explain
why the universe is nearly homogeneous on the horizon scale.

2) Isotropy problem. We need to understand why all directions in
the universe are similar to each  other, why there is no overall
rotation of the universe. etc.

3) Horizon problem. This one is closely related to the homogeneity
problem. If different parts of  the universe have not been in a
causal contact when the universe was born, why do they look so
similar?

4) Flatness problem. Why $\Omega \approx 1$? Why parallel lines do
not intersect?

5) Total entropy problem. The total entropy of the observable part
of the universe is  greater than $10^{87}$. Where did this huge
number come from? Note that the lifetime of a closed universe
filled with hot gas with total entropy $S$  is $S^{2/3}\times
10^{-43}$ seconds \cite{book}. Thus $S$ must be huge. Why?

6) Total mass problem. The total mass of the observable part of
the universe has mass  $\sim 10^{60} M_p$.  Note also that the
lifetime of a closed universe filled with nonrelativistic
particles of total mass $M$ is ${M\over M_P} \times 10^{-43}$
seconds. Thus $M$ must be huge. But why?

7) Structure formation problem. If we manage to explain the
homogeneity of the universe, how can  we explain the origin of
inhomogeneities required for the large scale structure formation?

8) Monopole problem, gravitino problem, etc.

This list is very long. That is why it was not easy to propose any
alternative to inflation even  before we learned that $\Omega
\approx 1$, $n_{s}\approx 1$, and that the perturbations responsible
for galaxy formation are mostly adiabatic, in agreement with the
predictions of the simplest inflationary models.

There were many attempts to propose an alternative to inflation in recent years. In general, this could be a very healthy tendency. If one of these attempts will succeed, it will be of great importance. If none of them are successful, it will be an additional demonstration of the advantages of inflationary cosmology. However, since the stakes are high, we are witnessing a growing number of premature announcements of success in developing an alternative cosmological theory. 

\subsubsection*{Cosmic strings and textures}

15 years ago the models of structure formation due to topological defects or textures were advertised in popular press as the models that ``match the explanatory 
triumphs of inflation while rectifying its major failings''         
 \cite{SperTur}. However, it was
clear from the very beginning that these theories at best could
solve only one problem (structure formation) out of 8 problems
mentioned above. The true question was not whether one
can replace inflation by the theory of cosmic strings/textures,
but whether inflation with cosmic strings/textures is better than
inflation without cosmic strings/textures. Recent observational
data favor the simplest version of inflationary theory, without
topological defects, or with an extremely small (few percent)
admixture of the effects due to cosmic strings.

\subsubsection*{Pre-big bang}

An attempt to avoid the use of the standard inflationary mechanism (though still use a stage of inflation prior to the big bang) was made in 
the pre-big bang scenario \cite{PBB}. This scenario is based on the assumption that eventually one will find a solution of the cosmological singularity problem
and learn how one could transfer small perturbations of the metric
through the singularity. This problem still remains unsolved, see e.g.  \cite{nogo}. Moreover, a detailed investigation of the homogeneity, isotropy and flatness problems in the pre-big bang scenario demonstrated that  the stage of the pre-big bang inflation introduced in \cite{PBB} is insufficient to solve the major cosmological problems  \cite{Kaloper:1998eg}.

\subsubsection*{Ekpyrotic/cyclic scenario}

A similar situation emerged with the introduction of the
ekpyrotic  scenario  \cite{KOST}. The original version of this
theory  claimed that this scenario can solve all
cosmological problems without using the stage of inflation, i.e.
without a prolonged stage of an accelerated expansion of the
universe, which was called in \cite{KOST} ``superluminal
expansion.'' However, the original ekpyrotic scenario contained many significant errors and did not work. It is sufficient to say that instead of the big bang expected in  \cite{KOST}, there was a big crunch  \cite{KKL,KKLTS}.

The ekpyrotic scenario was replaced by the cyclic scenario, which used an infinite number of periods of expansion and contraction of the universe  \cite{cyclic}. The origin of the required scalar field potential in this model remains unclear, and the very existence of the cycles postulated in \cite{cyclic} have not been demonstrated. When we analyzed this scenario  using the particular potential given in  \cite{cyclic}, and took into account the effect of particle production in the early universe, we found a very different cosmological regime \cite{Felder:2002jk,Linde:2002ws}.

The original version of the cyclic scenario relied on the existence of an infinite number of very long stages of ``superluminal expansion'', i.e.
inflation, in order to solve the major cosmological problems. In this sense, the original version of the cyclic scenario was not a true alternative to inflationary scenario, but its rather peculiar version. The main difference between the usual inflation and the cyclic inflation, just as in the case of topological defects and
textures, was the mechanism of generation of density perturbations.
However, since the theory of density perturbations in cyclic
inflation requires a solution of the cosmological singularity
problem, it is difficult to say anything definite about it. Originally there was a hope that the cosmological singularity problem will be solved in the context of string theory, but despite the attempts of the best experts in string theory, this problem remains unsolved  \cite{Liu:2002ft,Horowitz:2002mw,Berkooz}. 

Most of the authors believe that even if the singularity problem will be solved,  the spectrum of perturbations  in the standard version of this scenario involving only one scalar field after the singularity will be very non-flat. One may introduce more complicated versions of this scenario, involving many scalar fields. In this case, under certain assumptions about the way the universe passes through the singularity,  one may find a special regime where isocurvature perturbations in one of these fields are converted to adiabatic perturbations with a nearly flat spectrum. A recent discussion of  this scenario shows that this regime requires extreme fine-tuning of  initial conditions  \cite{Koyama:2007mg}. Moreover, the instability of the solutions in this regime, which was found in \cite{Koyama:2007mg}, implies that it may be very easy to switch from one regime to another under the influence of small perturbations. This may lead to domain structure of the universe and large perturbations of the metric \cite{Sasaki}. If this is the case, no fine-tuning of initial conditions will help.

One of the latest versions of the cyclic scenario attempted to avoid the long stage of  accelerated expansion (low-scale inflation) and  to make the universe homogeneous using some specific features of  the ekpyrotic collapse \cite{Erickson:2006wc}. The authors assumed that  the universe was homogeneous prior to its collapse on the scale that becomes greater  than the scale of the observable part of the universe during the next cycle. Under this assumption, they argued that the perturbations of the metric produced during each subsequent cycle do not interfere with the perturbations of the metric produced in the next cycle.  As a result,  {  if   the universe was homogeneous from the very beginning}, it remains homogeneous on the cosmologically interesting scales in all subsequent cycles. 

Is this a real solution of the homogeneity problem? The initial size of the  part of the universe, which is required to be homogeneous in this scenario prior to the collapse, was many orders of magnitude greater than the Planck scale. How homogeneous should it be? If we want the inhomogeneities to be produced due to amplification of quantum perturbations, then the initial classical perturbations of the field responsible for the isocurvature perturbations must be incredibly small, smaller than its quantum fluctuations. Otherwise the initial classical inhomogeneities of this field  will be amplified by the same processes that amplified its quantum fluctuations and  will dominate the spectrum of perturbations after the bounce \cite{KKL}. This  problem  is closely related to the problem mentioned above  \cite{Koyama:2007mg,Sasaki}.

Recently there was an attempt to revive the original (non-cyclic) version of the ekpyrotic scenario  by involving  a nonsingular bounce. This regime requires violation of the null energy condition \cite{KKLTS}, which usually leads to a catastrophic vacuum instability and/or causality violation. One may hope to avoid these problems in the ghost condensate theory \cite{Arkani-Hamed:2003uy}; see a series of recent papers on this subject \cite{Creminelli:2006xe,Buchbinder:2007ad,Creminelli:2007aq}. However, even the authors of the ghost condensate theory emphasize that a fully consistent version of this theory is yet to be constructed \cite{Adams:2006sv}, and that it may be incompatible with basic gravitational principles   \cite{Arkani-Hamed:2007ky}. 

In addition, just as  the ekpyrotic scenario with the singularity  \cite{Koyama:2007mg}, the new version of the ekpyrotic theory requires two fields, and a conversion of the isocurvature perturbations to adiabatic perturbations    \cite{Lehners:2007ac}. Once again, the initial state of the universe in this scenario must be extremely homogeneous: the initial classical perturbations of the field responsible for the isocurvature perturbations must be  smaller than its quantum fluctuations.  It does not seem possible to solve this problem without further extending this  exotic model and making it a part of an even more complicated scenario.

\subsubsection*{String gas scenario}

Another attempt to solve some of the cosmological problems without using inflation was made by Brandenberger {\it et al}  in the context of string gas cosmology \cite{Nayeri:2005ck,Brandenberger:2006xi}. The authors admitted that their model did not solve the flatness problem, so it was not a real alternative to inflation. However, they claimed that their model provided a   non-inflationary mechanism of production of perturbations of the metric with a flat spectrum.

It would be quite interesting and important to have a new mechanism of generation of perturbations of the metric based on string theory. Unfortunately, a detailed analysis of the scenario  proposed in \cite{Nayeri:2005ck,Brandenberger:2006xi} revealed that  some of its essential ingredients were either unproven or incorrect \cite{Kaloper:2006xw}. For example, the theory of generation of perturbations of the metric used in \cite{Nayeri:2005ck} was formulated in the Einstein frame, where the usual Einstein equations are valid. On the other hand, the  bounce and the string gas cosmology were described in string frame. Then both of these results were combined  without distinguishing between different frames and a proper translation from one frame to another.  

If one makes all calculations carefully (ignoring many other unsolved problems of this scenario), one finds that the perturbations generated in their scenario have blue spectrum with $n = 5$, which is ruled out by cosmological observations \cite{Kaloper:2006xw}. After the conference   ``Inflation + 25,'' where this issue was actively debated,  the authors of \cite{Nayeri:2005ck,Brandenberger:2006xi} issued two new papers reiterating their claims \cite{Nayeri:2006uy,Brandenberger:2006vv}, but eventually they agreed with our conclusion expressed at this conference: the spectrum of perturbations of the metric in this scenario is blue, with $n = 5$,  see Eq. (43) of  \cite{Brandenberger:2006pr}. This rules out the  models proposed in \cite{Nayeri:2005ck,Brandenberger:2006xi,Nayeri:2006uy,Brandenberger:2006vv}. Nevertheless, as often happens with various alternatives to inflation, some of  the authors of Refs.  \cite{Nayeri:2005ck,Brandenberger:2006xi,Nayeri:2006uy,Brandenberger:2006vv} still claim  that  their basic scenario  remains intact and propose its further modifications \cite{Brandenberger:2006pr,Biswas:2006bs,Brandenberger:2007qi}.

\subsubsection*{Mirage bounce}

Paradoxes with the choice of frames appear in other works on bounces in cosmology as well. For example, in \cite{Germani:2006pf} it was claimed that one can solve all cosmological problems in the context of  mirage cosmology. However, as  explained in \cite{Kachru:2002kx}, in the Einstein frame in this scenario the universe does not evolve at all.  

To clarify the situation without going to technical details, one may consider the following analogy. We know that all particles in our body get their masses due to spontaneous symmetry breaking in the standard model. Suppose that the Higgs field initially was out of the minimum of its potential, and experienced oscillations. During these oscillations the masses of electrons and protons also oscillated. If one measures the size of the universe in units of the (time-dependent) Compton wavelengths of the electron (which could seem to be a good idea), one would think that the scale factor of the universe oscillates (bounces) with the frequency equal to the Higgs boson mass. And yet, this ``cosmological evolution'' with bounces of the scale factor is  an illusion, which  disappears if one  measures the distances in units of the Planck length $M_{p}^{{-1}}$ (the Einstein frame).

In addition, the mechanism of generation of density perturbations used  in \cite{Germani:2006pf}   was borrowed from  the paper by Hollands and Wald \cite{Hollands:2002yb}, who suggested yet another alternative mechanism of generation of perturbations of the metric.  However, this mechanism requires investigation of thermal processes at the density 90 orders of magnitude greater than the Planck density, which makes all calculations unreliable \cite{Kofman:2002cj}.

\subsubsection*{Bounce in quantum cosmology}

Finally, I should mention Ref. \cite{Peter:2006hx}, where it was argued that under certain conditions one can have a bouncing universe and produce perturbations of the metric with a flat spectrum in the context of quantum cosmology. However, the model of Ref. \cite{Peter:2006hx} does not solve the flatness and homogeneity problems.  A more detailed analysis revealed that the wave function of the universe proposed in \cite{Peter:2006hx} makes the probability of a bounce of a large universe exponentially small  \cite{Peterprivite}. The authors are working on a modification of their model, which, as they hope, will not suffer from this problem. 

\

To conclude,  at the moment it is hard to see any real alternative to inflationary cosmology, despite an active search for such alternatives. All of the proposed alternatives are based on various attempts to solve the singularity problem: One should either construct a bouncing nonsingular cosmological solution, or learn what happens to the universe when it goes through the singularity. This problem  bothered cosmologists for nearly a century, so it would be great to find its solution, quite independently of the possibility to find an alternative to inflation. None of the proposed alternatives can be consistently formulated until this problem is solved. 

In this respect, inflationary theory has a very important advantage: it works practically independently of the solution of the singularity problem. It can work equally well after the singularity, or after the bounce, or after the quantum creation of the universe. This fact is especially clear in the  eternal inflation scenario: Eternal inflation makes the processes which occurred near the big bang practically irrelevant for the subsequent evolution of the universe.

\section{Naturalness of chaotic inflation}

Now we will return to the discussion of various versions of inflationary theory.
Most of them are based on the idea of
chaotic initial conditions, which is the trademark of the chaotic
inflation scenario. In the simplest versions of chaotic inflation
scenario with the potentials $V \sim \phi^n$, the process of
inflation occurs at $\phi>1$, in Planck units. Meanwhile, there
are many other models where inflation may occur at $\phi \ll 1$.

There are several reasons why this difference may be important.
First of all, some authors argue that  the generic expression for
the effective potential can be cast in the form
\begin{equation}\label{LythRiotto}
V(\phi) = V_0 +\alpha \phi+ {m^2\over 2} \phi^2 +{\beta\over 3}
\phi^3+ {\lambda\over 4} \phi^4 + \sum_n \lambda_n
{\phi^{4+n}\over {M_p}^n}\, ,
\end{equation}
and then they assume that generically $\lambda_n = O(1)$, see e.g.
Eq. (128) in \cite{LythRiotto}.  If this assumption were correct,
one would have little control over the behavior of $V(\phi)$ at
$\phi > M_p$.

Here we have written $M_p$ explicitly, to expose the implicit
assumption made in \cite{LythRiotto}.  Why do we write $M_p$ in
the denominator, instead of $1000 M_p$? An intuitive reason is
that quantum gravity is non-renormalizable, so one should
introduce a cut-off at momenta $k \sim M_p$. This is a reasonable
assumption, but it does not imply validity of Eq.
(\ref{LythRiotto}). Indeed, the constant part of the scalar field
appears in the gravitational diagrams not directly, but only via
its effective potential $V(\phi)$ and the masses of particles
 interacting with the scalar field $\phi$. As a result,
the terms induced by quantum gravity effects are suppressed not by
factors ${\phi^n \over {M_p}^n}$, but by factors  $V\over {M_p}^4$
and $m^2(\phi)\over {M_p}^2$ \cite{book}. Consequently, quantum
gravity corrections to $V(\phi)$ become large not at $\phi
> M_p$, as one could infer from (\ref{LythRiotto}), but only at
super-Planckian energy density, or for super-Planckian masses.
This justifies our use of the simplest chaotic inflation models.

The simplest way to understand this argument is to consider the
case where the potential of the field $\phi$ is a constant,
$V=V_0$. Then the theory has a {\it shift symmetry}, $\phi \to
\phi +c$. This symmetry is not broken by perturbative quantum
gravity corrections, so no such terms as $\sum_n \lambda_n
{\phi^{4+n}\over {M_p}^n}$ are generated. This symmetry may be
broken by nonperturbative quantum gravity effects (wormholes?
virtual black holes?), but such effects, even if they exist, can
be made exponentially small \cite{Kallosh:1995hi}.

On the other hand, one may still wonder whether there is any reason not to add the terms like  $\lambda_n
{\phi^{4+n}\over {M_p}^n}$ with $\lambda = O(1)$ to the theory. Here I will make a simple argument which may help to explain it. I am not sure whether this argument should   be taken too seriously, but I find it quite amusing and unexpected.

Let us consider a theory with the potential 
\begin{equation}\label{LythRiotto2}
V(\phi) = V_0 +\alpha \phi+ {m^2\over 2} \phi^2  + \lambda_n
{\phi^{4+n}\over {M_p}^n}\, +{\xi\over 2} R\phi^{2} \ .
\end{equation}
The last term is added to increase the generality of our discussion by considering fields non-minimally coupled to gravity, including the conformal fields with $\xi = 1/6$.

Suppose first that $m^{2} = \lambda_{n} = 0$. Then the theory can describe our ground state with a slowly changing vacuum energy only if $V_{0}+\alpha\phi < 10^{{-120}}$, $\alpha < 10^{{-120}}$ \cite{300}. This theory cannot describe inflation because $\alpha$ is too small to produce the required density perturbations.

Let us now  add the quadratic term. Without  loss of generality one can make a redefinition of the field $\phi$ and $V_{0}$ to remove the linear term:
\begin{equation}\label{LythRiotto3}
V(\phi) = V_0+ {m^2\over 2} \phi^2   \ .
\end{equation}
This is the simplest version of chaotic inflation. The maximal value of the field $\phi$ in this scenario is given by the condition ${m^2\over 2} \phi^2 \sim 1$ (Planckian density), so the maximal amount of inflation in this model is $\sim e^{\phi^{2}/4} \sim e^{1/m^{2}}$.

If, instead, we would consider a more general case with the three terms $ {m^2\over 2} \phi^2  + \lambda_n
{\phi^{4+n}\over {M_p}^n}\, +{\xi\over 2} R\phi^{2}$, the maximal amount of inflation would be 
\begin{equation}\label{LythRiotto4}
N < \exp \left [{\rm min} \{ m^{-2}, \lambda_{n}^{-2/n},\xi^{-1}\}\right]  \ .
\end{equation}
The last constraint appears because the effective gravitational constant becomes singular at $\phi^{2} \sim \xi^{-1}$.

Thus,  if any of the constants $\lambda_{n}^{2/n}$ or $\xi$ is greater than $m^{2}$, the total amount of inflation will be exponentially smaller than in the simplest theory ${m^2\over 2} \phi^2$. Therefore one could argue that if one has a possibility to choose between different inflationary theories, as in the string theory landscape, then the largest fraction of the volume of the universe will be in the parts of the multiverse with $\lambda_{n}^{2/n}, \xi \ll m^{2}$.  One can easily check that for $\lambda_{n}^{2/n}, \xi \lesssim m^{2}$ the higher order terms can be ignored at the last stages of inflation, where $\phi = O(1)$. In other words, the theory behaves as purely quadratic during the last stages of inflation when the observable part of the universe was formed. 

One can come to the same conclusion if one takes into account  only the part of inflation at smaller values of the field $\phi$, when the stage of eternal inflation is over. This suggests that the simplest version of chaotic inflation scenario is  the best.

Of course, this is just an argument. Our main goal here was not to promote the model ${m^2\over 2} \phi^2 $, but to demonstrate that the  considerations of naturalness (e.g. an assumption that all $\lambda_{n}$ should be large) depend quite crucially on the underlying assumptions. In the example given above, a very simple change of these assumptions (the emphasis on the total volume of the post-inflationary universe) was sufficient to explain naturalness of the simplest model  ${m^2\over 2} \phi^2 $. However, the situation may become quite different if instead of the simplest theory of a scalar field combined with general relativity one starts to investigate more complicated models,  such  as supergravity and string theory.

\section{Chaotic inflation in supergravity}

In the simplest models of inflation, the field $\phi$ itself does not have any direct physical meaning; everything depends only on its functions such as the masses of particles and the scalar potential. However, in more complicated theories the scalar field $\phi$ itself may have physical (geometric) meaning,  which may constrain the
possible values of the fields during inflation. The most important
example is given by $N = 1$ supergravity.

The F-term potential of the complex scalar field $\Phi$ in
supergravity is given by the well-known  expression (in units $M_p
= 1$):
\begin{equation}\label{superpot}
V = e^{K} \left[K_{\Phi\bar\Phi}^{-1}\, |D_\Phi W|^2
-3|W|^2\right].
\end{equation}
Here $W(\Phi)$ is the superpotential, $\Phi$ denotes the scalar
component of the superfield  $\Phi$; $D_\Phi W= {\partial W\over
\partial \Phi} + {\partial K\over \partial \Phi} W$. The kinetic
term of the scalar field is given by $K_{\Phi\bar\Phi}\,
\partial_\mu \Phi \partial _\mu \bar\Phi$. The standard textbook
choice of the K\"ahler potential corresponding to the canonically
normalized fields $\Phi$ and $\bar\Phi$ is $K = \Phi\bar\Phi$, so
that $K_{\Phi\bar\Phi}=1$.

This immediately reveals a problem: At $\Phi > 1$ the potential is
extremely steep.  It blows up as $e^{|\Phi|^2}$, which makes it
very difficult to realize chaotic inflation in supergravity at
$\phi \equiv \sqrt 2|\Phi| > 1$. Moreover, the problem persists
even at small $\phi$. If, for example, one considers the simplest
case when there are many other scalar fields  in the theory  and
the superpotential does not depend on the inflaton field $\phi$,
then Eq. (\ref{superpot}) implies that at $\phi \ll 1$ the
effective mass of the inflaton field is $m^2_\phi = 3H^2$. This
violates the  condition $m^2_\phi \ll H^2$ required for successful
slow-roll inflation (so-called $\eta$-problem).

The major progress in SUGRA inflation during the last decade was
achieved in the context of the models of the hybrid inflation
type, where inflation may occur at $\phi \ll 1$. Among the best
models are the F-term inflation, where different contributions to
the effective mass term $m^2_\phi$ cancel \cite{F}, and D-term
inflation \cite{D}, where the dangerous term $e^K$ does not affect
the potential in the inflaton direction. A detailed discussion of
various versions of hybrid inflation in supersymmetric theories
can be found in \cite{LythRiotto,pterm,Binetruy:2004hh,Lyth:2007qh,Kallosh:2007ig}.

However, hybrid inflation occurs only on a relatively small energy
scale, and many of its versions do not lead to eternal inflation.
Therefore it would be nice to obtain inflation in a context of a
more general class of supergravity models.

This goal seemed very difficult to achieve; it took almost 20
years to find a natural realization of chaotic inflation model in
supergravity. Kawasaki, Yamaguchi and Yanagida suggested to take
the K\"ahler potential
\begin{equation} K = {1\over 2}(\Phi+\bar\Phi)^2
+X\bar X \end{equation}
 of the fields $\Phi$ and $X$, with the
superpotential $m\Phi X$ \cite{jap}.

At the first glance, this K\"ahler potential may seem somewhat
unusual. However, it can be obtained from the standard K\"ahler
potential $K =  \Phi \bar\Phi  +X\bar X$  by adding terms
$\Phi^2/2+\bar\Phi^2/2$, which do not give any contribution to the
kinetic term of the scalar fields $K_{\Phi\bar\Phi}\,
\partial_\mu \Phi \partial _\mu \bar\Phi$. In other words, the new
K\"ahler potential, just as the old one, leads to canonical
kinetic terms for the fields $\Phi$ and $X$, so it is as simple
and legitimate as the standard textbook K\"ahler potential.
However, instead of the U(1) symmetry with respect to rotation of
the field $\Phi$ in the complex plane, the new K\"ahler potential
has a {\it shift symmetry}; it does not depend on the imaginary
part of the field $\Phi$. The shift symmetry is broken only by the
superpotential.

This leads to a profound  change of the potential
(\ref{superpot}): the dangerous term $e^K$ continues growing
exponentially in the direction $(\Phi +\bar\Phi)$, but it remains
constant in the direction $(\Phi - \bar \Phi )$. Decomposing the
complex field $\Phi$ into two real scalar fields, $ \Phi = {1\over
\sqrt 2} (\eta +i\phi)$, one can find the resulting potential
$V(\phi,\eta,X)$ for $\eta, |X| \ll 1$:
\begin{equation}\label{superpot1}
V = {m^2\over 2} \phi^2 (1 + \eta^2) + m^2|X|^2.
\end{equation}
This potential has a deep valley, with a minimum at $\eta = X =0$. At $\eta, |X| > 1$ the potential grows up exponentially.
Therefore the fields $\eta$ and $X$ rapidly fall down towards
$\eta = X =0$, after which the potential for the field $\phi$
becomes $V = {m^2\over 2} \phi^2$. This provides  a very simple
realization of eternal chaotic inflation scenario in supergravity
\cite{jap}. This model can be extended to include theories with
different power-law potentials, or models where inflation begins
as in the simplest versions of chaotic inflation scenario, but
ends as in new or hybrid inflation, see e.g.
\cite{Yamaguchi:2001pw,Yok}.

The existence of shift symmetry was also the basis of the natural inflation scenario \cite{natural}. The basic assumption of   this scenario was that the axion field in the first approximation is massless because the flatness of the axion direction is protected by $U(1)$ symmetry. Nonperturbative corrections lead to the axion potential $V(\phi) = V_{0} (1+\cos(\phi/f_{a}))$.  If the `radius' of the axion potential $f_{a}$ is sufficiently large, $f_{a} \gtrsim 3$, inflation near the top of the potential becomes possible. For much greater values of $f_{a}$ one can have inflation near the minimum of the axion potential, where the potential is quadratic \cite{Savage:2006tr}.

The natural inflation scenario was proposed  back in 1990, but until now all attempts to realize this scenario in supergravity failed. First of all, it was difficult to find theories with large $f_{a}$. More importantly, it was difficult to stabilize the radial part of the axion field. A possible model of natural inflation in supergravity was constructed only very recently   \cite{Kallosh:2007ig}.

Unfortunately, we still do not know how one could incorporate the models discussed in this section to string theory. We will briefly describe some features of inflation in string theory, and refer the readers to a more detailed presentation in  \cite{Kallosh:2007ig}.

\section{Towards Inflation in String Theory}
\subsection{de Sitter vacua in string theory}

For a long time, it seemed rather difficult to obtain inflation in
M/string theory. The main problem here was the stability of
compactification of internal dimensions. For example, ignoring
non-perturbative effects to be discussed below, a typical
effective potential of the effective 4d theory obtained by
compactification in string theory of type IIB can be represented
in the following form:
\begin{equation}
V(\varphi,\rho,\phi) \sim e^{\sqrt 2\varphi -\sqrt6\rho}\ \tilde
V(\phi)
\end{equation}
Here $\varphi$ and $\rho$ are canonically normalized fields
representing the dilaton field and the volume of the compactified
space; $\phi$ stays for all other fields, including the inflaton field.

If $\varphi$ and $\rho$ were constant, then the potential $\tilde
V(\phi)$ could drive inflation.  However, this does not happen
because of the steep exponent $e^{\sqrt 2\varphi -\sqrt6\rho}$,
which rapidly pushes the dilaton field $\varphi$ to $-\infty$, and
the volume modulus $\rho$ to $+\infty$. As a result, the radius of
compactification becomes infinite; instead of inflating, 4d space
decompactifies and becomes 10d.

Thus in order to describe inflation one should first learn how to
stabilize the dilaton and the volume modulus. The dilaton
stabilization was achieved in \cite{GKP}. The most difficult
problem was to stabilize the volume. The solution of this problem
was found in \cite{KKLT} (KKLT construction). It consists of two
steps.

\begin{figure}[h!]
\centering\leavevmode\epsfysize=5.5cm \epsfbox{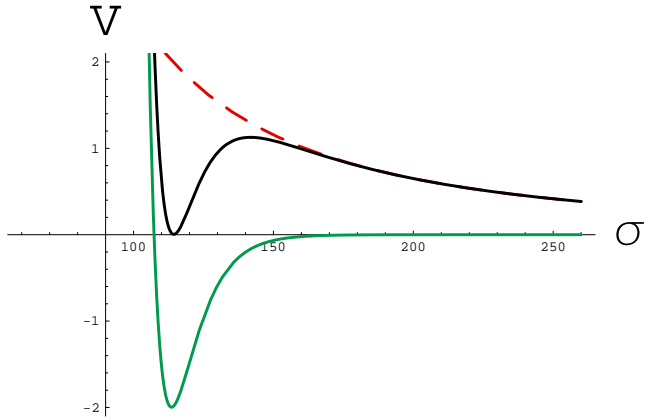} \caption[fig1]
{KKLT potential as a function of $\sigma = {\rm Re}\,\rho$. Thin green line corresponds to AdS stabilized potential for $W_0 =- 10^{{-4}}$,
$A=1$, $a =0.1$. Dashed line shows the additional term,
which appears either due to the contribution of a $\overline{D3}$ brane or of a
D7 brane. Thick black line shows the resulting potential  with a very small but positive value of $V$ in the minimum. The potential is shown multiplied by $10^{15}$.} \label{1}
\end{figure}

First of all, due to a combination of effects related to warped
geometry of the compactified space and nonperturbative effects
calculated directly in 4d (instead of being obtained by
compactification), it was possible to obtain a supersymmetric AdS
minimum of the effective potential for $\rho$. In the original version of the KKLT scenario, it was done in the theory with the  K\"ahler potential
\begin{equation}
K = -3\log (\rho+\bar\rho)
, \end{equation} and with the nonperturbative superpotential of the form \begin{equation}\label{KKLTsp}
W=W_0+
Ae^{-a\rho}, 
\end{equation} 
with $a = 2\pi/N$.
The corresponding effective potential for the complex field $\rho =
\sigma +i\alpha$ had a minimum at finite, moderately large values of the volume modulus field $\sigma_{0}$, which  fixed the
volume modulus  in a state with a negative vacuum energy. Then
 an anti-${D3}$ brane with the positive energy $\sim
\sigma^{-2}$ was added. This addition uplifted the minimum of the potential to
the state with a positive vacuum energy, see Fig. \ref{1}.

Instead of adding an anti-${D3}$ brane, which explicitly breaks
supersymmetry, one can add  a D7 brane with fluxes. This results
in the appearance of a D-term which has a similar dependence on
$\rho$, but leads to spontaneous supersymmetry breaking
\cite{Burgess:2003ic}. In either case, one ends up with a
metastable dS state which can decay by tunneling and formation of
bubbles of 10d space with vanishing vacuum energy density. The
decay rate is extremely small \cite{KKLT}, so for all practical
purposes, one obtains an exponentially expanding de Sitter space
with the stabilized volume of the internal space\footnote{It is
also possible to find de Sitter solutions in noncritical string
theory \cite{Str}.}.

\subsection{Inflation in string theory}

There are two different versions of string inflation. In the first version, which we will call modular inflation, the inflaton field is  associated with one of the moduli, the scalar fields which are already present in the KKLT construction. In the second version, the inflaton is related to the distance between branes moving in the compactified space. (This scenario should not be confused with  inflation in the brane world scenario  \cite{Arkani-Hamed:1998rs,Randall:1999ee}. This is a separate interesting subject, which we are not going to discuss in this paper.)

\subsubsection {Modular inflation}

\begin{figure}[hbt]
\centering\leavevmode\epsfysize=7.5cm \epsfbox{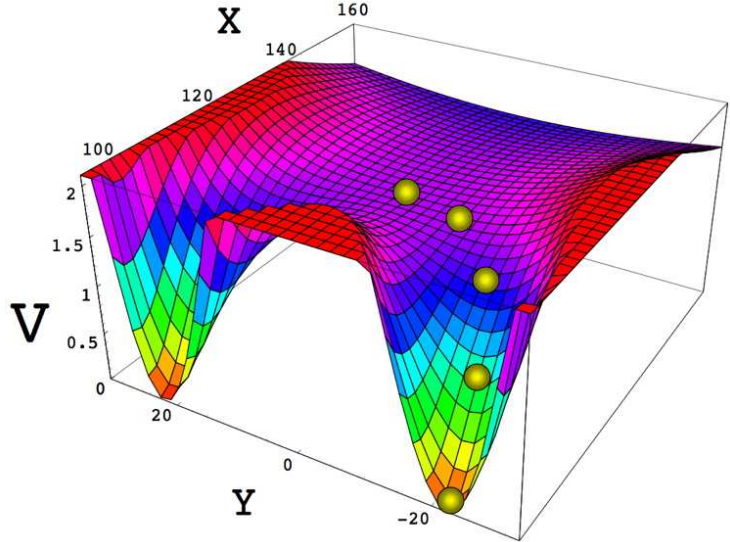}

\caption[fig1] {Plot for the  potential in the racetrack model (rescaled by
$10^{16}$). Here X stays for $\sigma = {\rm Re }\, \rho$ and Y stays for $\alpha = {\rm Im }\, \rho$. Inflation begins in a vicinity of the saddle point at $X_{\rm saddle}=123.22$, $ Y_{\rm saddle}=0$.
Units are $M_p=1$.\label{F1}}
\end{figure}

An example of the  KKLT-based  modular inflation is provided by the racetrack inflation model of Ref. \cite{Blanco-Pillado:2004ns}. It uses a slightly more complicated superpotential  
\begin{equation}\label{race}
W=W_0+
Ae^{-a\rho} + B e^{-b\rho}.
 \end{equation}
The potential of this theory has a saddle point as a function of  the real and the complex part of the volume modulus: It has a local minimum in the direction $\rm Re\, \rho$,  which is simultaneously a very flat maximum with respect to $\rm Im\,\rho$. Inflation occurs during a slow rolling of the field $\rm Im\,\rho$ away from this maximum (i.e. from the saddle point). The existence of this regime requires a significant fine-tuning of  parameters of the superpotential. However, in the context of the string landscape scenario describing from $10^{100}$ to $10^{1000}$ different vacua (see below), this may not be such a big issue. A nice feature of this model is that it does not require adding any new branes  to the original KKLT scenario, i.e. it is rather economical. 

Other interesting models of moduli inflation were developed in \cite{Conlon:2005jm,Lalak:2005hr,Blanco-Pillado:2006he,Bond:2006nc}. An interesting property of all of these models is the existence of the regime of eternal slow-roll inflation. This property distinguishes modular inflation form the brane inflation scenario to be discussed below.

\subsubsection {Brane inflation}

During the last few years there were many suggestions how to
obtain hybrid inflation in string theory by considering motion of
branes in the compactified space, see \cite{Dvali:1998pa,Quevedo}
and references therein. The main problem of all of these models
was the absence of stabilization of the compactified space. Once
this problem was solved for dS space \cite{KKLT}, one could try to
revisit these models and develop models of brane inflation
compatible with the volume stabilization.

The first idea \cite{KKLMMT} was to consider a pair of D3 and
anti-D3 branes in the warped geometry studied in \cite{KKLT}. The
role of the inflaton field  $\phi$ in this model, which is known as the KKLMMT model, 
could be played by the interbrane
separation. A description of this situation in terms of the
effective 4d supergravity involved K\"ahler potential
\begin{equation}
K = -3\log (\rho+\bar\rho -k(\phi,\bar\phi)), \end{equation}
 where the function
$k(\phi,\bar\phi)$ for the inflaton field $\phi$, at small $\phi$,
was taken in the simplest form $k(\phi,\bar\phi)= \phi\bar\phi$.
 If one makes  the simplest
assumption that the superpotential does not depend on $\phi$, then
the $\phi$ dependence of the potential (\ref{superpot})  comes
from the term $e^K =(\rho+\bar\rho - \phi\bar\phi)^{-3}$.
Expanding this term near the  stabilization point $\rho = \rho_0$,
one finds that the inflaton field has a mass $m^2_\phi = 2H^2$.
Just like the similar relation $m^2_\phi = 3H^2$ in the simplest
models of supergravity, this is not what we want for inflation.

One way to solve this problem is to consider $\phi$-dependent
superpotentials. By doing so, one may fine-tune $m^2_\phi$ to be
$O(10^{-2}) H^2$ in a vicinity of the point where inflation occurs
\cite{KKLMMT}. Whereas fine-tuning is certainly undesirable, in
the context of string cosmology it may not be a serious drawback.
Indeed, if there exist many realizations of string theory (see Section \ref{land}), then one might argue that all realizations not
leading to inflation can be discarded, because they do not
describe a universe in which we could live. This makes the issue
of fine-tuning less problematic. 
Inflation in the KKLMMT model and its generalizations was studied by many authors; see  \cite{Kallosh:2007ig} and references therein.

Can we avoid fine-tuning altogether? One of the possible ideas is
to find theories with some kind of shift symmetry. Another
possibility is to construct something like D-term inflation, where
the flatness of the potential is not spoiled by the term $e^K$.
Both of these ideas were combined together in Ref.
\cite{Hsu:2003cy} based on the model of D3/D7 inflation in string
theory \cite{renata}. In this model the K\"ahler potential is
given by
\begin{equation}
K = -3\log (\rho+\bar\rho) -{1\over 2}(\phi - \bar \phi)^2,
\end{equation}
and superpotential depends only on $\rho$. The role of the inflaton field is played by the field $s = {\rm Re}\, \phi$, which represents the distance between the D3 and D7 branes. The shift symmetry
$s \to s+c$ in this model is related to the requirement of
unbroken supersymmetry of branes in a BPS state.

The effective potential with respect to the field $\rho$ in this
model coincides with the KKLT potential
\cite{KKLT,Burgess:2003ic}. The
potential is exactly flat in the direction of the inflaton field $s$, until one adds a hypermultiplet of  other fields $\phi_{\pm}$,  which break
this flatness due to quantum corrections and produce a logarithmic potential for the field $s$. The resulting potential with respect to the fields $s$ and $\phi_{\pm}$ is very similar to the potential of D-term hybrid inflation \cite{D}.

\begin{figure}[h!]
\centering\leavevmode\epsfysize=6.5cm \epsfbox{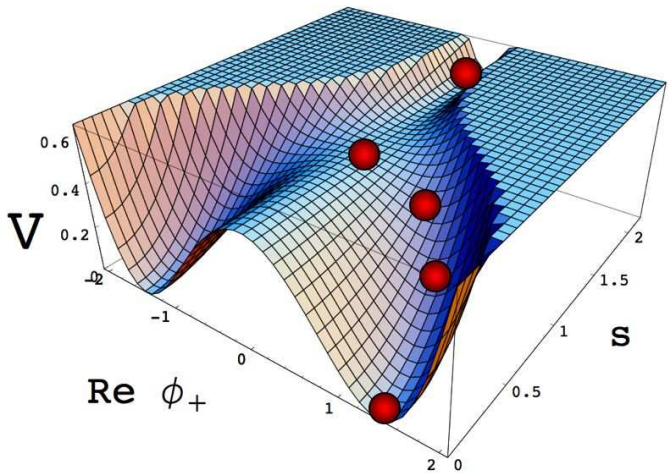} \caption[fig2]
{Inflationary potential as a function of the inflaton field $s$ and
${\rm Re}\, \phi_+$. In the beginning, the field $s$ rolls
along the valley $\phi_+=0$, and then it falls down to the KKLT minimum.}
\label{4}
\end{figure}

During inflation, $\phi_{\pm}=0$, and the field $s$ slowly rolls down to its smaller values. When it becomes sufficiently small, the theory becomes unstable with respect to generation of the field $\phi_{+}$, see Fig. \ref{4}.  The fields $s$ and  $\phi_{+}$ roll down to the KKLT minimum, and inflation ends.
For the latest developments in D3/D7 inflation see \cite{Dasgupta:2004dw,Chen:2005ae}.

All inflationary models discussed above were formulated in the context of Type IIB string theory with the KKLT stabilization. A discussion of the possibility to obtain inflation in the heterotic string theory with stable compactification can be found in  \cite{Buchbinder:2004nt,Becker:2005sg}.

Finally, we should mention that making the effective potential flat is not the only way to achieve inflation. There are some models with nontrivial kinetic terms where inflation may occur even without any potential \cite{kinfl}. One may also consider models with steep potentials but with anomalously large kinetic terms for the scalar fields  see e.g. \cite{Dim}. In application to string theory,  such models, called `DBI inflation,' were developed in \cite{Silverstein:2003hf}.

In contrast to the moduli inflation, none of the existing versions of the brane inflation  allow the slow-roll eternal inflation  \cite{Chen:2006hs}.

\section{Scale of inflation, the gravitino mass, and the amplitude of the gravitational waves}

\begin{figure}[h!]
\centering\leavevmode\epsfysize=5.5cm \epsfbox{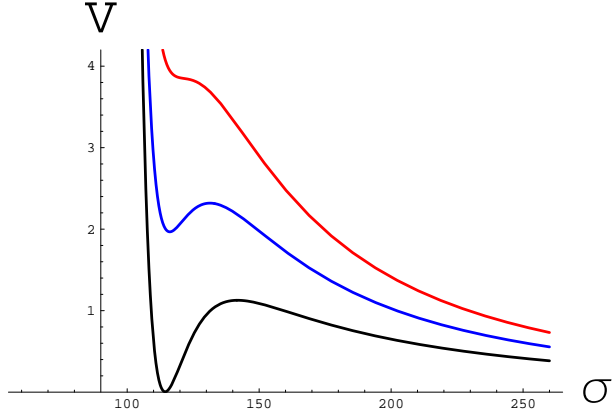} \caption[fig2]
{The lowest curve with dS minimum is the one from the KKLT model. The height of the barrier in this potential is of the order $m_{3/2}^{2}$.  The second line shows the $\sigma$-dependence of the inflaton potential. When one adds it to the theory, it always appears divided by $\sigma^{n}$, where in the simplest cases $n = 2$  or 3.  Therefore an addition of the inflationary potential lifts up the potential at small $\sigma$. The top curve shows that when the  inflation potential becomes too large, the barrier disappears, and the internal space decompactifies. This explains the origin of the constraint $H\lesssim   m_{3/2}$.  } \label{2}
\end{figure}

So far, we did not discuss relation of the new class of models with particle phenomenology. This relation is rather unexpected and may impose strong constraints on particle phenomenology and on inflationary models:  in the simplest models based on the KKLT mechanism the Hubble constant $H$ and the inflaton mass $m_{\phi}$   are  smaller than the gravitino mass  \cite{Kallosh:2004yh},
\begin{equation}
 m_{\phi} \ll H \lesssim m_{{3/2}} \ .
\end{equation}
The reason for the constraint $H \lesssim m_{{3/2}}$ is that the height of the barrier stabilizing the KKLT minimum is $O(m_{3/2}^{2})$. Adding a large vacuum energy density  to the KKLT potential, which is required for inflation, may destabilize it, see Fig. \ref{2}. The constraint $ m_{\phi} \ll H$ is a consequence of the slow-roll conditions.

Therefore if one believes in the standard SUSY phenomenology with $m_{{3/2}} \lesssim O(1)$ TeV, one should find a realistic particle physics model where  inflation occurs  at a density at least 30 orders of magnitude below the Planck energy density. Such models are possible, but their parameters should be substantially different from the parameters used in all presently existing models of string theory inflation.

An interesting observational consequence of this result is that the amplitude of the gravitational waves in all string inflation models of this type should be extremely small. Indeed, according to Eq. (\ref{eq:rvh}), one has 
$ {r} \approx 3\times  10^{7}~V  \approx 10^{8}~ H^{2}$, which implies that 
\begin{equation}\label{bound}
r \lesssim 10^{8}~m_{{3/2}}^{2}  \ ,
\end{equation}
in Planck units. In particular, for $m_{{3/2}} \lesssim 1$ TeV $\sim 4 \times 10^{-16}~ M_{p}$, which is in the range most often discussed by SUSY phenomenology, one has \cite{kl2007}
\begin{equation}
r \lesssim 10^{-24} \ .
\end{equation}
If CMB experiments find that $r \gtrsim 10^{-2}$, then this will imply, in the class of theories described above, that 
\begin{equation}
m_{{3/2}} \gtrsim 10^{-5}~ M_{p} \sim 2.4 \times 10^{13}~{\rm GeV}  \ ,
\end{equation}
which is 10 orders of magnitude greater than the standard gravitino mass range discussed by particle phenomenologists.

There are several different ways to address this problem. First of all, one may try to construct realistic particle physics models with superheavy gravitinos \cite{DeWolfe:2002nn,Arkani-Hamed:2004fb}. 

\begin{figure}[h!]
\centering\leavevmode\epsfysize=5cm \epsfbox{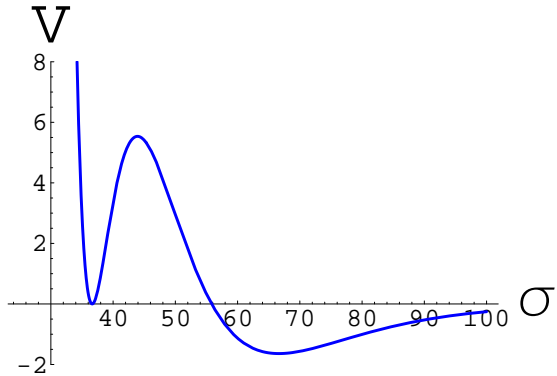} \caption[fig2]
{The potential in the theory (\ref{race}) for $A=1,\ B=-5,\ a=2\pi/100,\ b=2\pi/50,\ W_0= -0.05$. A Minkowski minimum at $V=0$ stabilizes the volume at $\sigma_{0}\approx 37$.  The height of the barrier in this model is not correlated with the gravitino mass, which   vanishes  if the system is trapped in Minkowski vacuum. Therefore in this model one can avoid the constraint $H \lesssim m_{{3/2}}$   \cite{Kallosh:2004yh}. } \label{3a}
\end{figure}

Another possibility is to consider models with the racetrack superpotential containing at least two exponents (\ref{race}) and find such parameters that the supersymmetric minimum of the potential even before the uplifting  occurs at zero energy density   \cite{Kallosh:2004yh}, which would mean $m_{3/2} = 0$, see Fig. \ref{3a}. Then, by a slight change of parameters one can get the gravitino mass squared much smaller than the height of the barrier, which removes the constraint $H \lesssim m_{{3/2}}$.

Note, however, that in order to have  $H^{2} \sim V \sim 10^{-10}$ with $m_{{3/2}} \lesssim 1$ TeV $\sim 4 \times 10^{-16}~ M_{p}$ in the model of  \cite{Kallosh:2004yh} one would need to fine-tune the parameters of the theory with an incredible precision. This observation further strengthens the results  of  \cite{Baumann:2006cd,Bean:2007hc}, which imply that the tensor perturbations produced in all known versions of string theory inflation are undetectably small.

One could argue that since the existing versions of string theory inflation predict tensor modes with an extremely small amplitude, there is no sense to even try to detect them. From our perspective, however, the attitude should be opposite. There is a class of  inflationary models that predict $r$ in the range from $0.3$ to $10^{-2}$, see Section \ref{observ}, so it makes a lot of sense to test this range of $r$ even though the corresponding models have not been constructed as yet in the context of string theory. 

If the tensor modes will be found, the resulting situation will be similar to the situation with the discovery of the acceleration of the universe. This discovery initially puzzled string theorists, since none of the  versions of string theory which existed 5 years ago could describe an accelerating universe in a stable vacuum state with a positive energy density. Eventually this problem was resolved with the development of the KKLT construction.

A possible discovery of tensor modes could lead to another constructive crisis since it may  rule out many existing versions of string inflation and string phenomenology, and it may imply that the gravitino must be superheavy. Thus, investigation of gravitational waves produced during inflation may serve as a unique source of information about string theory and fundamental physics in general \cite{kl2007}.

\section{Initial conditions for the low-scale inflation and topology of the universe}\label{torus}

One of the advantages of the simplest versions of the chaotic
inflation scenario is that inflation may begin in the universe immediately after its creation at the largest
possible energy density $M_p^{4}$, of a smallest possible size (Planck
length), with the smallest possible mass $M \sim M_p$ and with the
smallest possible entropy $S = O(1)$. This provides a true
solution to the flatness, horizon, homogeneity, mass and entropy
problems \cite{book}. 

Meanwhile, in the new inflation scenario (more accurately, in the hilltop version of the chaotic inflation scenario), inflation
occurs on the mass scale 3 orders of magnitude below $M_p$, when
the total size of the universe was very large. If, for example, the universe is closed, its total mass  at the beginning of new inflation
must be  greater than $10^6 M_p$, and its total entropy must be
greater than $10^9$. In other words, in order to explain why the
entropy of the universe at present is greater than $10^{87}$ one should
assume that it was extremely large from the very beginning. Then it becomes difficult to understand  why such a large universe was homogeneous. This
does not look like a real solution of the  problem of initial conditions. 

Thus one may wonder whether  it  possible to solve the problem of initial conditions for the low scale inflation?
The answer to this question is positive though perhaps somewhat unexpected: The simplest way to solve the problem of initial conditions for the low scale inflation is to consider a compact flat  or open universe with nontrivial topology (usual flat or open universes are infinite). The universe may initially look like a nearly homogeneous torus of a Planckian size containing just one or two photons or gravitons. It can be shown that such a universe continues expanding and remains homogeneous until the onset of inflation, even if inflation occurs only on a very low energy scale \cite{ZelStar,chaotmix,topol4,Coule,Linde:2004nz}. 

Consider, e.g. a flat compact universe having the topology of a
torus, $S_1^3$,
\begin{equation}\label{2t}
ds^2 = dt^2 -a_i^2(t)\,dx_i^2
\end{equation}
with identification $x_i+1 = x_i$ for each of the three dimensions. Suppose
for  simplicity that $a_1 = a_2 = a_3 = a(t)$. In this case the curvature
of the universe and the Einstein equations written in terms of $a(t)$ will be
the same as in the infinite flat Friedmann universe with metric $ds^2 = dt^2
-a^2(t)\,d{\bf x^2}$. In our notation, the scale factor $a(t)$ is equal to the
size of the universe in Planck units $M_{p}^{{-1}} = 1$.

Let us assume,  that at the
Planck time $t_p \sim M_p^{-1}=1$ the universe was radiation dominated, $V\ll
 T^4 = O(1)$. Let us also assume that at the Planck time the total size of the
box was Planckian, $a(t_p) = O(1)$. In such case the whole universe initially
contained only $O(1)$ relativistic particles such as photons or gravitons, so
that the total entropy of the whole universe was O(1).

 The size of the universe dominated by relativistic particles was growing as
$a(t) \sim \sqrt t$, whereas the mean free path of the gravitons was growing as
$H^{-1}\sim t$. If the initial size of the universe was $O(1)$, then at the
time  $t \gg 1$ each particle (or a gravitational perturbation of metric)
within one cosmological time would run all over the torus many times, appearing
in all of its parts with nearly equal probability. This effect, called
``chaotic mixing,'' should lead to a rapid homogenization of the universe
\cite{chaotmix,topol4}. Note, that to achieve a modest degree of homogeneity
required for inflation to start when the density of ordinary matter drops down,
we do not even need chaotic mixing. Indeed, density perturbations do not grow
in a universe dominated by ultrarelativistic particles if  the size of the
universe is smaller than $H^{-1}$. This is exactly what
happens in our model. Therefore the universe should remain relatively
homogeneous until the thermal energy density drops below $V$ and inflation
begins. And once it happens, the universe rapidly becomes very homogeneous.

Thus we see that in this scenario, just as in the simplest chaotic inflation scenario, inflation begins if we had a sufficiently homogeneous domain of a smallest possible size (Planck scale), with the smallest possible mass (Planck mass), and with the total entropy O(1). The only additional requirement is that this domain should have identified sides, in order to make a flat or open universe compact. We see no reason to expect that the probability of formation of such domains is strongly suppressed.

One can come to a similar conclusion from a completely different point of view. Investigation of the quantum creation of a closed or  an infinite open inflationary universe with $V\ll 1$ shows that this process is forbidden at the classical level, and therefore it occurs only due to tunneling. As a result, the probability of this process is exponentially suppressed \cite{Linde:1983mx,Vilenkin:1984wp,Open}. Meanwhile,  creation of the flat or open universe  is possible without any need for the tunneling, and therefore there is no exponential suppression for the probability of quantum creation of a topologically nontrivial compact flat or open inflationary universe \cite{ZelStar,Coule,Linde:2004nz}.

These results suggest that if inflation can occur only  much below the Planck density, then the compact topologically nontrivial flat or open universes should be much more probable than the standard Friedmann universes described in every textbook on cosmology. This possibility is quite natural in the context of string theory, where all internal dimensions are supposed to be compact. Note, however, that if the stage of inflation  is sufficiently long, it should make the observable part of the universe so large that its topology does not affect observational data. 

The problem of initial conditions in string cosmology has several other interesting features. The most important one is the existence of an enormously large number of metastable de Sitter vacuum states, which makes the stage of exponential expansion of the universe almost inevitable. We will discuss this issue in the next section.

\section{Inflationary multiverse,  string theory landscape and the anthropic principle}\label{land}

For many decades people have tried to explain strange correlations
between the properties of our universe, the masses of elementary
particles, their coupling constants, and the fact of our existence.
We know that we could not live in a 5-dimensional universe, or in a
universe where the electromagnetic coupling constant, or the masses
of  electrons and protons would be just a few times greater or
smaller than their present values. These and other similar
observations have formed the basis for the anthropic principle.
However, for a long time many scientists  believed that the universe
was given to us as a single copy, and therefore speculations
about these magic coincidences could not have any scientific
meaning. Moreover, it would require a wild stretch of imagination and a certain degree of arrogance to assume that somebody was creating one universe after another, changing their parameters and fine-tuning their design, doing all of that for the sole purpose of making the universe suitable for our existence.

The situation changed dramatically with the invention of
inflationary cosmology. It was realized that inflation may divide
our universe into many exponentially large domains corresponding to
different metastable vacuum states, forming a huge inflationary
multiverse \cite{linde1982,nuff,Eternal}. The total number of
such vacuum states in string theory can be enormously large, in the range of $10^{100}$ or $10^{1000}$
\cite{Lerche:1986cx,Bousso:2000xa,KKLT,Douglas}. A combination of these
two facts  gave rise to what the experts in inflation call `the inflationary multiverse,' \cite{book,LLM,Linde:2002gj} and string theorists call  
`the string theory landscape'  \cite{Susskind:2003kw}. 

This leads to an interesting twist in the theory of initial conditions. Let us assume first that we live in one of the many metastable de Sitter minima, say, $dS_{i}$. Eventually this dS state decays, and each of the {\it points} belonging to this initial state  jumps to another vacuum state, which may have either a smaller vacuum energy, or a greater vacuum energy (transitions of the second type are possible because of the gravitational effects). But if the decay probability   is not too large, then the total {\it volume} of the universe remaining in the  state $dS_{i}$  continues growing exponentially \cite{Guth:1982pn}. This is eternal inflation of the old inflation type. If the bubbles of the new phase correspond to another de Sitter space, $dS_{j}$, then some parts of the space $dS_{j}$ may jump back to the   state  $dS_{i}$. On the other hand, if the tunneling goes to a Minkowski vacuum, such as the uncompactified 10d vacuum corresponding to the state with $\sigma \to \infty$ in   Fig. \ref{1}, the subsequent jumps to dS states no longer occur. Similarly, if the tunneling goes to the state with a negative vacuum energy, such as the AdS vacuum  in Fig. \ref{3a}, the interior of the bubble of the new vacuum rapidly collapses. Minkowski and AdS vacua of such type are called terminal vacua, or sinks.

If initial conditions in a certain part of the universe are such that it goes directly go to the sink, without an intermediate stage of inflation, then it will never return back, we will be unable to live there, so for all practical purposes such initial conditions (or such parts of the universe) can be discarded (ignoring for the moment the possibility of the resurrection of the universe after the collapse). On the other hand, if some other part of the universe goes to one of the dS states, the process of eternal inflation begins, which eventually produces an inflationary multiverse consisting of all possible dS states. This suggests that all initial conditions that allow life as we know it to exist, inevitably lead to formation of an eternal inflationary multiverse.

This scenario assumes that the vacuum transitions may bring us from any part of the string theory landscape to any other part. Here we should note that the theory of such transitions accompanied by the change of fluxes was developed for the case where dS states are not stabilized \cite{Brown:1987dd,Bousso:2000xa}. A generalization of this theory for the string landscape scenario based on the KKLT mechanism of vacuum stabilization is rather nontrivial. As of now,  the theory of such transitions was fully developed only for the transitions where the scalar fields change but the fluxes remain unchanged \cite{Ceresole:2006iq}. It might happen that the landscape is  divided into separate totally disconnected islands, but this does not seem likely \cite{Clifton:2007en}. Even if the landscape is not fully transversable,  one may probe all parts of the inflationary multiverse by considering the wave function of the universe corresponding to the possibility of its quantum creation in the states with different values of fluxes \cite{Linde:1984ir,Hawking:2006ur}.

The string theory landscape describes an incredibly large set of {\it discrete} parameters. However, the theory of inflationary multiverse goes even further. Some of the features of our world are determined not by the final values of the fields in the minima of their potential in the landscape, but by the dynamical, time-dependent values, which these fields were taking at different stages of the evolution of the inflationary universe. This introduces a large set of {\it continuous} parameters, which may take different values in different parts of the universe. For example, in the theory of dark energy, inflationary fluctuations may divide the universe into exponentially large parts with the effective value of the cosmological constant  taking a continuous range of values \cite{300}. In such models, the effective cosmological constant $\Lambda$ becomes a continuous parameter.  Similarly, inflationary fluctuations of the axion field make  the density of  dark matter  a continuous parameter, which takes  different values in different parts of the universe \cite{Linde:1987b,Rees}. Another example of a continuous parameter is the baryon asymmetry $n_{b}/n_{\gamma}$, which can  take different values in different parts of the universe in the Affleck-Dine scenario of baryogenesis \cite{Affleck:1984fy,Linde:gh}.

This means that the same physical theory may yield exponentially large  parts of the universe that have diverse properties.  This provided the first scientific justification of the anthropic principle: we find  ourselves inside a  part of the universe with our kind of physical laws not because the parts with different properties are impossible or improbable, but simply because we cannot live there \cite{linde1982,nuff}. 
 
This fact can help us understand many otherwise mysterious features of our world. The simplest example concerns the dimensionality of our universe. String theorists usually assume  that the universe is 10, or 11 dimensional, so why do we live in the universe where only 4  dimensions of space-time are large? There were many attempts to address this question, but no convincing answer was found. This question became even more urgent after the development of the KKLT construction. Now we know that all de Sitter states, including the state in which we live now, are either unstable or   metastable. They tend to decay by producing bubbles of a collapsing space, or of a 10-dimensional Minkowski space. So what is wrong about the 10-dimensional universe if it is so naturally appears in string theory?

The answer to this question was given in 1917 by Paul Ehrenfest  \cite{ehr}: In space-time with dimensionality $d > 4$, gravitational forces between distant
bodies fall off faster than $r^{-2}$, and in space-time with $d<4$, the general theory of relativity tells us that such forces are absent altogether.
This rules out the existence of stable planetary systems for $d\not = 4$. 
A similar conclusion is valid for atoms:  stable atomic systems could not exist for $d> 4$. This means that we do not need to prove that the 4d space-time is a {\it necessary} outcome of string cosmology  (in fact, it does not seem to be the case). Instead of that, we only need to make sure that the 4d space-time is {\it possible}.

Anthropic considerations may help us to understand  why the amount of dark matter is approximately 5 times greater than the amount of normal matter \cite{Linde:1987b,Rees} and why the baryon asymmetry is so small, $n_{b}/n_{\gamma} \sim 10^{{-10}}$  \cite{Linde:gh}. But perhaps the most famous example of this type is related to the cosmological constant problem. 

Naively, one could expect vacuum energy to be equal to the Planck density,  $\rho_{\Lambda} \sim 1$, whereas the recent observational data show that $\rho_{\Lambda} \sim 10^{-120}$, in Planck units, which is approximately 3 times greater than the density of other matter in the universe. Why is it so small but nonzero? Why $\rho_{\Lambda}$ constitutes is about 3 times greater than the density of other types of matter in the universe now? Note that long ago the density of matter was much greater than $\rho_{\Lambda}$, and in the future it will be much smaller.

The first anthropic solution to the cosmological constant problem in the context of inflationary cosmology was proposed  in 1984
\cite{Linde:1984ir}. The basic assumption was that the vacuum energy density is a sum of the scalar field potential $V(\phi)$ and the energy of fluxes $V(F)$. According to  \cite{Linde:1983mx}, quantum creation of the universe is not suppressed if the universe is created at the Planck energy density, $V(\phi) +V(F) = O(1)$, in Planck units. Eventually the field $\phi$ rolls to its minimum at some value $\phi_0$, and the vacuum energy becomes $\Lambda = V(\phi_0) + V(F)$. Since initially $V(\phi)$ and $V(F)$  could take any values with nearly equal probability, under the condition $V(\phi) +V(F) = O(1)$, we get a flat probability distribution to find a universe with a given value of the cosmological constant after inflation, $\Lambda = V(\phi_0) + V(F)$, for $\Lambda \ll 1$. The flatness of this probability distribution is crucial, because it allows us to study the probability of emergence of life for different $\Lambda$. Finally, it was  argued in   \cite{Linde:1984ir}  that life as we know it is possible only for  $|\Lambda|  \lesssim \rho_{0}$, where $\rho_{0} \sim 10^{{-120}}$ is the present energy density of the universe.  This fact, in combination with inflation, which makes such universes exponentially large, provided a possible solution of the cosmological constant problem.

Shortly after that, several other anthropic solutions to the cosmological constant problem were  proposed  \cite{Banks}. All of them were based on the assumption that life as we know it is possible only for  $-\rho_{0} \lesssim \rho_{\Lambda} \lesssim \rho_{0}$. This bound seemed almost self-evident to many of us at that time, and therefore in \cite{Linde:1984ir,Banks} we concentrated on the development of the theoretical framework where the anthropic arguments could be applied  to the cosmological constant.

The fact that $\rho_{\Lambda}$ could not be much smaller than $-\rho_{0}$ was indeed quite obvious, since such a universe would rapidly collapse. However, the origin of the constraint $\rho_{\Lambda} \lesssim \rho_{0}$ was much less trivial. The first attempt to justify it was made in 1987 in the famous paper by Weinberg \cite{Weinberg:1987dv}, but the constraint obtained there allowed the cosmological constant to be three orders of magnitude greater than its present value.

Since that time, the anthropic approach to the cosmological constant problem developed in two different directions. First of all, it became possible, under certain assumptions, to significantly strengthen the constraint on the positive cosmological constant, see e.g. \cite{Weinberg1998,Garriga:1999hu,Tegmark:2005dy,Lineweaver:2007qh}.  The final result of these investigations, $|\Lambda|  \lesssim O(10)\ \rho_{0} \sim 10^{-119}$, is very similar to the bound used in \cite{Linde:1984ir}.

Simultaneously, new models have been developed which may allow us to put anthropic approach to the cosmological constant problem on a firm ground. In particular, the existence of a huge number of vacuum states in string theory  implies that in different parts of our universe, or in its different quantum states, the cosmological constant may take all of its possible values, from $-1$ to $+1$, with an increment which may be as small as $10^{-1000}$. If  the prior probability to be in each of these vacua does not depend strongly on $\Lambda$, one can justify the anthropic bound on $\Lambda$ using the methods of \cite{Weinberg1998,Garriga:1999hu,Tegmark:2005dy,Bousso:2007kq,Lineweaver:2007qh}.

However, the issue of probabilities in eternal inflation is  very delicate, so one should approach anthropic arguments with some care. 
For example, one may try to calculate the probability to be born in a part of the universe with given properties {\it at a given point}. One can do this using comoving coordinates,  which are not expanding during inflation \cite{Starobinsky:1986fx,Goncharov:1987ir,Garriga:1997ef,Bousso:2006ev,Clifton:2007en,Podolsky:2007vg}. However, it is not obvious whether the calculation of the probabilities of physical processes at a given point, ignoring expansion of the universe, should be used in anthropic considerations. Most of the physical entities which could be associated with ``points'' did not even exist before the beginning of inflation: protons did not exist, photons did not exist, galaxies did not exist. They appeared only after inflation, and their total number, and the total number of observers, is proportional to the growth of volume during inflation. 

This leads to the volume-weighted \cite{LLM,Bellido,Mediocr,Vilenkin:2006qf}, or pocket-weighted \cite{Vilenkin:2006qf,GDVW,Vilenkin:2006qg}   probability measures \cite{volume}. The main problem with this approach is the embarrassment of riches: The total volume of the universe occupied by any particular vacuum state, integrated over the indefinitely long history of the eternally inflating universe, is infinitely large. Thus we need to compare infinities, which is a very ambiguous task, with the answer depending on the choice of the cut-off procedure.

The volume-weighted probability measure proposed in  \cite{LLM} is based on the calculation of  the  ratio of the volumes of the parts of the universe with different properties. This is possible because if we wait long enough, eternal inflation  approaches a stationary regime. Different parts of the universe expand and transform to each other. As a result, the total volume of all parts of the universe of each particular type grow at the same rate, and the ratio of their volumes becomes time-independent \cite{LLM}. 

This method is very good for describing the map of the  inflationary multiverse, but  in order to use it in anthropic considerations one should make some additional steps. According to \cite{Bellido}, instead of calculating the ratio of volumes in different  vacuum states at different  densities and temperatures, we should calculate the total volume of {\it new} parts of the universe where life becomes possible. This ratio is related to the incoming probability current through the hypersurface of the end of inflation, or the hypersurface of a fixed density or temperature. If one uses the probability measure of \cite{LLM} for anthropic considerations (which was {\it not} proposed in \cite{LLM}), one may encounter the so-called youngness paradox \cite{Guth:2007ng,Tegmark:2004qd}. If one uses the prescription of \cite{Bellido}, this paradox does not appear  \cite{Linde:2006nw}.

The results of the calculations by this method are very sensitive to the choice of the   time parametrization \cite{Bellido,Linde:2006nw}. However, a recent investigation of this issue indicates that it may be possible to resolve this problem \cite{lindemeasure}. The main idea is that the parts of the universe with different properties approach the stationary regime of eternal inflation at different times. This fact was not taken into account in our earlier papers \cite{LLM,Bellido}; the calculations of the probabilities started everywhere at the same time, even if the corresponding parts of the universe did not yet approach the stationary regime. If we start comparing the volumes of different part of the universe not at the same time after the beginning of inflation, but at the same time since the beginning of  the stationarity regime,  the dependence on the time parametrization disappears, at least in the simple cases where we could verify this property   \cite{lindemeasure}. 

As we already mentioned, there are many other proposals for the calculations of probabilities in inflationary multiverse, see e.g.  \cite{Vilenkin:2006qf,Vilenkin:2006qg}. The results of some of these methods are not sensitive to the choice of time parametrization, but  they do depend on the choice of the  cut-off. A detailed discussion of this series of proposals can be found in  \cite{GDVW,Aguirre:2006ak,Winitzki:2006rn}.

While discussing all of these approaches one should keep in mind yet another possibility: It is quite possible that it does not make much sense to compare infinities and talk about the probability of events that already happened. Instead of doing it, one should simply study our part of the universe, take these data as an initial input for all subsequent calculations, and study  conditional probabilities for the quantities which we did not measure yet \cite{Linde:2006nw}. This is a standard approach used by experimentalists who continuously re-evaluate the probability of various outcomes of their future experiments on the basis of other experimental data. The non-standard part is that we should be allowed to use {\it all} of our observations,  including our knowledge of our own properties,  for the calculation of conditional probabilities. 

Let us apply this limited approach to the cosmological constant problem. 20 years
ago, we already knew that our life is carbon-based, and that the amplitude of density perturbations required for the formation of galaxies was about $10^{{-5}}$. We did not know yet what was the vacuum energy, and the prevailing
idea was that we did not have much choice anyway. But with the
discovery of inflation, we learned that the universe could be
created differently, with different values of the cosmological
constant in each of  its parts
created by eternal inflation. This allowed us to propose several
different anthropic solutions to the cosmological constant problem
based on the assumption that,
for the given value of the amplitude of density perturbations and other already measured parameters, {\it we} cannot live in a universe with $|\Lambda| \gg 10^{-120}$. 
If observations would show that the cosmological constant  were a million times
smaller than the anthropic bound, then we would be surprised, and a
theoretical explanation of this anomaly would be in order. As of
now, the small value of the cosmological constant does not look too
surprising, so for a while we can concentrate on solving many other problems which cannot be addressed by anthropic considerations.

Within  this approach, one should not  vary the constants of nature that were already known at the time when the predictions were made. In doing so, one is at risk of repeating the old argument that the bomb does not hit the same spot twice: It   is correct  only until the first hit, after which the probabilities should be re-evaluated. Similarly, one should not omit the word `anthropic' from the `anthropic principle' and should not replace investigation of the probability of {\it our} life with the study of  life in general: We are trying to explain {\it our} observations rather than than the possible observations made by some abstract information-processing devices. This can help us to avoid some paradoxes recently discussed in the literature  \cite{Garriga:2005ee,Harnik:2006vj,Starkman:2006at}.

From this discussion   it should be clear that we do not really know yet which of the recently developed approaches to the  theory of the inflationary multiverse is going to be more fruitful, and how far we will be able to go in this direction.  One  way or another, it would be very difficult  to forget about what we just learned and return to our search for the theory which unambiguously explains all parameters of our world. Now we know that some features of our part of the universe may have an unambiguous explanation, whereas some others can be purely environmental and closely correlated with our own existence.

When inflationary theory was first proposed, its main goal was to address many problems which at that time could seem rather metaphysical: Why is our universe so big? Why is it so uniform? Why parallel lines do not intersect? It took some time before we got used to the idea that the large size, flatness and uniformity of the universe should not be dismissed as trivial facts of life. Instead of that, they should be considered  as  observational data requiring an explanation. 

Similarly,  the existence of an amazingly strong correlation between our own properties and the values of many parameters of our world, such as the masses and charges of electron and proton, the value of the gravitational constant, 
the amplitude of spontaneous symmetry breaking in the electroweak theory, 
the value of the vacuum energy, and the dimensionality of our world, is an experimental fact requiring an explanation. A combination of the theory of inflationary multiverse and the string theory landscape provide us with a unique framework where this explanation can possibly be found.  

\section{Conclusions}
 
25 years ago, inflationary theory looked like an exotic product of vivid scientific imagination. Some of us believed that it possesses such a great explanatory potential that it must be correct; some others thought that it is too good to be true. Not many expected that it is possible to verify any of its predictions in our lifetime.  Thanks to the enthusiastic work of many scientists, inflationary theory is gradually becoming a widely accepted cosmological paradigm, with many of its predictions being confirmed by observational data.  

However, while the basic principles of inflationary cosmology are rather well established, many of its details are still changing with each new change of the theory of all fundamental interactions. Investigation of the  inflationary multiverse and the string theory landscape force us to think about problems which sometimes go beyond the well established boundaries of physics. This makes our life difficult, sometimes quite frustrating, but also very interesting, which is perhaps  the best thing that one could expect from the branch of science we were trying to develop during the last quarter of a century.

\section*{Acknowledgments}
 
I am grateful to the organizers of the Inflation + 25 conference and workshop for 
their hospitality. I would like to thank  my numerous collaborators who made my work on inflation so enjoyable, especially my old friends and frequent collaborators Renata Kallosh, Lev Kofman, and Slava Mukhanov.  This work was supported by the NSF grant 0244728 and by the Humboldt award. 

\


\begin{thebibliography}{99}

\bibitem{Linde:1974at}
  A.~D.~Linde,
``Is the Cosmological Constant Really a Constant?''
  JETP Lett.\  {\bf 19} (1974) 183
  [Pisma Zh.\ Eksp.\ Teor.\ Fiz.\  {\bf 19} (1974) 320].
  
\bibitem{Kirzhnits:1972ut}
D.~A.~Kirzhnits,
``Weinberg Model in the Hot Universe,''
  JETP Lett.\  {\bf 15} (1972) 529
  [Pisma Zh.\ Eksp.\ Teor.\ Fiz.\  {\bf 15} (1972) 745];
  D.~A.~Kirzhnits and A.~D.~Linde,
``Macroscopic Consequences Of The Weinberg Model,''
  Phys.\ Lett.\  B {\bf 42} (1972) 471.
  
\bibitem{Kirzhnits:1976ts}
  D.~A.~Kirzhnits and A.~D.~Linde,
``Symmetry Behavior In Gauge Theories,''
  Annals Phys.\  {\bf 101}, 195 (1976).
  
\bibitem{Linde:1978px}
  A.~D.~Linde,
``Phase Transitions In Gauge Theories And Cosmology,''
  Rept.\ Prog.\ Phys.\  {\bf 42}, 389 (1979).

\bibitem{Star} A.~A.~Starobinsky,
``Spectrum Of Relict Gravitational Radiation And The Early State
Of The  Universe,'' JETP Lett.\  {\bf 30}, 682 (1979) [Pisma Zh.\
Eksp.\ Teor.\ Fiz.\  {\bf 30}, 719 (1979)]; A.~A.~Starobinsky, ``A
New Type Of Isotropic Cosmological Models Without Singularity,''
Phys.\ Lett.\ B {\bf 91}, 99 (1980).


\bibitem{Mukh} V.~F.~Mukhanov and G.~V.~Chibisov,
``Quantum Fluctuation And `Nonsingular' Universe,'' JETP Lett.\
{\bf 33}, 532 (1981) [Pisma Zh.\ Eksp.\ Teor.\ Fiz.\  {\bf 33},
549 (1981)].




\bibitem{Guth} A.~H.~Guth,
``The Inflationary Universe: A Possible Solution To The Horizon
And Flatness Problems,'' Phys.\ Rev.\ D {\bf 23}, 347 (1981).



\bibitem{Hawking:1982ga}
  S.~W.~Hawking, I.~G.~Moss and J.~M.~Stewart,
``Bubble Collisions In The Very Early Universe,''
  Phys.\ Rev.\ D {\bf 26}, 2681 (1982).

\bibitem{Guth:1982pn}
  A.~H.~Guth and E.~J.~Weinberg,
``Could The Universe Have Recovered From A Slow First Order Phase
Transition?,''
  Nucl.\ Phys.\ B {\bf 212}, 321 (1983).
  

\bibitem{New}
A.~D.~Linde, ``A New Inflationary Universe Scenario: A Possible
Solution Of The Horizon, Flatness, Homogeneity, Isotropy And
Primordial Monopole Problems,'' Phys.\ Lett.\ B {\bf 108}, 389
(1982); A.~D.~Linde,
``Coleman-Weinberg Theory And A New Inflationary Universe Scenario,''
  Phys.\ Lett.\ B {\bf 114}, 431 (1982);
A.~D.~Linde,
``Temperature Dependence Of Coupling Constants And The Phase Transition In
The Coleman-Weinberg Theory,''
  Phys.\ Lett.\ B {\bf 116}, 340 (1982);
A.~D.~Linde,
``Scalar Field Fluctuations In Expanding Universe And The New Inflationary
Universe Scenario,''
  Phys.\ Lett.\ B {\bf 116}, 335 (1982). 

\bibitem{New2} A.~Albrecht and P.~J.~Steinhardt, ``Cosmology For Grand
Unified Theories With Radiatively Induced Symmetry Breaking,''
Phys.\ Rev.\ Lett.\  {\bf 48}, 1220 (1982).







\bibitem{Hawk} S.~W.~Hawking,
``The Development Of  Irregularities  In  A Single Bubble
Inflationary Universe,'' Phys.\ Lett.\ B {\bf 115}, 295 (1982);
A.~A.~Starobinsky, ``Dynamics Of Phase Transition In The New
Inflationary Universe Scenario And Generation Of Perturbations,''
Phys.\ Lett.\ B {\bf 117}, 175 (1982); A.~H.~Guth and S.~Y.~Pi,
``Fluctuations In The New Inflationary Universe,'' Phys.\ Rev.\
Lett.\  {\bf 49}, 1110 (1982); J.~M.~Bardeen, P.~J.~Steinhardt and
M.~S.~Turner, ``Spontaneous Creation Of Almost Scale - Free
Density Perturbations In An Inflationary Universe,'' Phys.\ Rev.\
D {\bf 28}, 679 (1983).

\bibitem{Mukh2} V.~F.~Mukhanov,
``Gravitational Instability Of The Universe Filled With A Scalar
Field,'' JETP Lett.\  {\bf 41}, 493 (1985) [Pisma Zh.\ Eksp.\
Teor.\ Fiz.\  {\bf 41}, 402 (1985)];  V. F. Mukhanov, {\it Physical Foundations of Cosmology}, Cambridge University Press, 2005.



\bibitem{book} A.D. Linde,  {\it  Particle  Physics  and
Inflationary Cosmology} (Harwood, Chur, Switzerland, 1990) [arXiv:hep-th/0503203].



\bibitem{Chaot} A.~D.~Linde,
``Chaotic Inflation,'' Phys.\ Lett.\ B {\bf 129}, 177 (1983).

\bibitem{Linde:1985ub}
  A.~D.~Linde,
``Initial Conditions For Inflation,''
  Phys.\ Lett.\  B {\bf 162} (1985) 281.
  
\bibitem{Linde:1983mx}
A.~D.~Linde, ``Quantum Creation Of The Inflationary Universe,'' Lett.\ Nuovo
Cim.\  {\bf 39}, 401 (1984).

\bibitem{Vilenkin:1984wp}
A.~Vilenkin, ``Quantum Creation Of Universes,'' Phys.\ Rev.\ D {\bf 30}, 509
(1984); A.~Vilenkin, ``Quantum Cosmology And The Initial State Of The Universe,''
Phys.\ Rev.\ D {\bf 37}, 888 (1988); A.~Vilenkin, ``The Interpretation Of The
Wave Function Of The Universe,'' Phys.\ Rev.\ D {\bf 39}, 1116 (1989).

\bibitem{Hartle:1983ai}
  J.~B.~Hartle and S.~W.~Hawking,
``Wave Function Of The Universe,''
  Phys.\ Rev.\  D {\bf 28}, 2960 (1983).
  
  
\bibitem{Open}
A.~D.~Linde, ``Quantum creation of an open inflationary universe,'' Phys.\
Rev.\ D {\bf 58}, 083514 (1998) [arXiv:gr-qc/9802038].

\bibitem{Linde:2006nw}
  A.~Linde,
``Sinks in the Landscape, Boltzmann Brains, and the Cosmological Constant Problem,''
  JCAP {\bf 0701}, 022 (2007)
  [arXiv:hep-th/0611043].
  



\bibitem{Gibbons:2006pa}
  G.~W.~Gibbons and N.~Turok,
``The measure problem in cosmology,''
  arXiv:hep-th/0609095.
  
\bibitem{Kofman:2002cj}
  L.~Kofman, A.~Linde and V.~F.~Mukhanov,
``Inflationary theory and alternative cosmology,''
  JHEP {\bf 0210}, 057 (2002)
  [arXiv:hep-th/0206088].
  
  \bibitem{koflingibbtur} L. Kofman and A. Linde, in preparation.


\bibitem{oldtheory}
A.~D.~Dolgov and A.~D.~Linde, ``Baryon Asymmetry In Inflationary
Universe,'' Phys.\ Lett.\ B {\bf 116}, 329 (1982); L.~F.~Abbott,
E.~Farhi and M.~B.~Wise, ``Particle Production In The New
Inflationary Cosmology,'' Phys.\ Lett.\ B {\bf 117}, 29 (1982).

\bibitem{KLS} L.~Kofman, A.~D.~Linde and A.~A.~Starobinsky,
``Reheating after inflation,'' Phys.\ Rev.\ Lett.\  {\bf 73}, 3195
(1994) [arXiv:hep-th/9405187]; L.~Kofman, A.~D.~Linde and
A.~A.~Starobinsky, ``Towards the theory of reheating after
inflation,'' Phys.\ Rev.\ D {\bf 56}, 3258 (1997)
[arXiv:hep-ph/9704452].


\bibitem{tach}  G.~N.~Felder, J.~Garcia-Bellido, P.~B.~Greene,
L.~Kofman, A.~D.~Linde and I.~Tkachev, ``Dynamics of symmetry
breaking and tachyonic preheating,'' Phys.\ Rev.\ Lett.\  {\bf
87}, 011601 (2001) [arXiv:hep-ph/0012142]; G.~N.~Felder, L.~Kofman
and A.~D.~Linde, ``Tachyonic instability and dynamics of
spontaneous symmetry breaking,'' Phys.\ Rev.\ D {\bf 64}, 123517
(2001) [arXiv:hep-th/0106179].

\bibitem{Desroche:2005yt}
  M.~Desroche, G.~N.~Felder, J.~M.~Kratochvil and A.~Linde,
``Preheating in new inflation,''
  Phys.\ Rev.\  D {\bf 71}, 103516 (2005)
  [arXiv:hep-th/0501080].
  
  \bibitem{latticeold} S.~Y.~Khlebnikov and I.~I.~Tkachev,
``Classical decay of inflaton,''
Phys.\ Rev.\ Lett.\  {\bf 77}, 219 (1996)
[arXiv:hep-ph/9603378]; S.~Y.~Khlebnikov and I.~I.~Tkachev,
``Resonant decay of Bose condensates,''
Phys.\ Rev.\ Lett.\  {\bf 79}, 1607 (1997)
[arXiv:hep-ph/9610477].

\bibitem{latticeeasy}
G.~N.~Felder and I.~Tkachev,
``LATTICEEASY: A program for lattice simulations of scalar fields in an
expanding universe,''
arXiv:hep-ph/0011159.

\bibitem{thermalization} G.~N.~Felder and L.~Kofman,
``The development of equilibrium after preheating,''
Phys.\ Rev.\ D {\bf 63}, 103503 (2001)
[arXiv:hep-ph/0011160];
R.~Micha and I.~I.~Tkachev,
``Turbulent thermalization,''
Phys.\ Rev.\ D {\bf 70}, 043538 (2004)
[arXiv:hep-ph/0403101];
 D.~I.~Podolsky, G.~N.~Felder, L.~Kofman and M.~Peloso,
``Equation of state and beginning of thermalization after preheating,''
  Phys.\ Rev.\  D {\bf 73}, 023501 (2006)
  [arXiv:hep-ph/0507096];
  G.~N.~Felder and L.~Kofman,
``Nonlinear inflaton fragmentation after preheating,''
  Phys.\ Rev.\  D {\bf 75}, 043518 (2007)
  [arXiv:hep-ph/0606256].

 
\bibitem{OvrStein}
B.~A.~Ovrut and P.~J.~Steinhardt, ``Supersymmetry And Inflation: A
New Approach,'' Phys.\ Lett.\ B {\bf 133}, 161 (1983); B.~A.~Ovrut
and P.~J.~Steinhardt, ``Inflationary Cosmology And The Mass
Hierarchy In Locally Supersymmetric Theories,'' Phys.\ Rev.\
Lett.\  {\bf 53}, 732 (1984); B.~A.~Ovrut and P.~J.~Steinhardt,
``Locally Supersymmetric Cosmology And The Gauge Hierarchy,''
Phys.\ Rev.\ D {\bf 30}, 2061 (1984); B.~A.~Ovrut and
P.~J.~Steinhardt, ``Supersymmetric Inflation, Baryon Asymmetry And
The Gravitino Problem,'' Phys.\ Lett.\ B {\bf 147}, 263 (1984).

\bibitem{Linde:cd}
A.~D.~Linde, ``Primordial Inflation Without Primordial
Monopoles,'' Phys.\ Lett.\ B {\bf 132} (1983) 317.

\bibitem{Boubekeur:2005zm}
L.~Boubekeur and D.~H.~Lyth,
``Hilltop inflation,''
  JCAP {\bf 0507}, 010 (2005)
  [arXiv:hep-ph/0502047].

\bibitem{Hybrid}
A.~D.~Linde, ``Axions in inflationary cosmology,'' Phys.\ Lett.\ B
{\bf 259}, 38 (1991); A.~D.~Linde, ``Hybrid inflation,'' Phys.\
Rev.\ D {\bf 49}, 748 (1994) [astro-ph/9307002].

\bibitem{Vilenkin:wt}
A.~Vilenkin and L.~H.~Ford, ``Gravitational Effects Upon
Cosmological Phase Transitions,'' Phys.\ Rev.\ D {\bf 26}, 1231
(1982).


\bibitem{Linde:uu}
A.~D.~Linde, ``Scalar Field Fluctuations In Expanding Universe And
The New Inflationary Universe Scenario,'' Phys.\ Lett.\ B {\bf
116}, 335 (1982).



\bibitem{Gross:1973id}
  D.~J.~Gross and F.~Wilczek,
``Ultraviolet Behavior Of Non-Abelian Gauge Theories,''
  Phys.\ Rev.\ Lett.\  {\bf 30}, 1343 (1973).
  
  \bibitem{Politzer:1973fx}
  H.~D.~Politzer,
``Reliable Perturbative Results For Strong Interactions?,''
  Phys.\ Rev.\ Lett.\  {\bf 30}, 1346 (1973).


\bibitem{LL}
A.R. Liddle and D. H. Lyth, {\it Cosmological Inflation and Large-Scale Structure}, (Cambridge University Press, Cambridge 2000).

\bibitem{WMAP}
  H.~V.~Peiris {\it et al.},
``First year Wilkinson Microwave Anisotropy Probe (WMAP) observations:
Implications for inflation,''
  Astrophys.\ J.\ Suppl.\  {\bf 148}, 213 (2003)
  [arXiv:astro-ph/0302225].
  
\bibitem{Martin:2006rs}
J.~Martin,
``Inflationary Perturbations: the Cosmological Schwinger Effect,''
  arXiv:0704.3540 [hep-th].


\bibitem{Tegmark}
M.~Tegmark {\it et al.},
``Cosmological Constraints from the SDSS Luminous Red Galaxies,''
  Phys.\ Rev.\  D {\bf 74}, 123507 (2006)
  [arXiv:astro-ph/0608632]. 
  

  
\bibitem{Kuo:2006ya}
  C.~L.~Kuo {\it et al.},
``Improved Measurements of the CMB Power Spectrum with ACBAR,''
  arXiv:astro-ph/0611198.
  
  
  
  
  \bibitem{Ax} A.~D.~Linde, ``Generation Of Isothermal Density Perturbations
In The Inflationary Universe,''  JETP Lett.\ {\bf40}, 1333 (1984) 
[Pisma Zh.\ Eksp.\ Teor.\ Fiz.\ {\bf 40}, 496 (1984)]; A.~D.~Linde,
``Generation Of Isothermal Density Perturbations In The Inflationary
Universe,''  Phys.\ Lett.\ B {\bf 158}, 375 (1985); L.~A.~Kofman,  ``What
Initial Perturbations May Be Generated In Inflationary Cosmological 
Models,''  Phys.\ Lett.\ B \textbf{173}, 400 (1986); L.~A.~Kofman and
A.~D.~Linde, ``Generation Of Density Perturbations In The Inflationary
Cosmology,''  Nucl.\ Phys.\ B \textbf{282}, 555 (1987); A.~D.~Linde and
D.~H.~Lyth,  ``Axionic Domain Wall Production During Inflation,''  Phys.\
Lett.\ B \textbf{246}, 353 (1990).

 

\bibitem{Mollerach:1989hu}  S.~Mollerach, ``Isocurvature Baryon
Perturbations And Inflation,''  Phys.\ Rev.\ D \textbf{42}, 313 (1990).

\bibitem{LM}  A.~D.~Linde and V.~Mukhanov,  ``Nongaussian isocurvature
perturbations from inflation,''  Phys.\ Rev.\ D \textbf{56}, 535 (1997) 
[arXiv:astro-ph/9610219]; A.~Linde and V.~Mukhanov,
``The curvaton web,''
  JCAP {\bf 0604}, 009 (2006)
  [arXiv:astro-ph/0511736].
  
  \bibitem{Enqvist:2001zp} K.~Enqvist and M.~S.~Sloth,
``Adiabatic CMB perturbations in pre big bang string cosmology,''
  Nucl.\ Phys.\  B {\bf 626}, 395 (2002)
  [arXiv:hep-ph/0109214].


\bibitem{LW}  D.~H.~Lyth and D.~Wands, ``Generating the curvature
perturbation without an inflaton,''  Phys.\ Lett.\ B {\bf 524}, 5 (2002) 
[arXiv:hep-ph/0110002].

\bibitem{Moroi:2001ct}  T.~Moroi and T.~Takahashi, ``Effects of cosmological
moduli fields on cosmic microwave background,''  Phys.\ Lett.\ B {\bf 522}, 215 (2001)  [Erratum-ibid.\ B {\bf 539}, 303 (2002)] 
[arXiv:hep-ph/0110096].


\bibitem{mod1} G.~Dvali, A.~Gruzinov and M.~Zaldarriaga, ``A new mechanism
for generating density perturbations from inflation,'' Phys.\ Rev.\ D 
{\bf 69}, 023505 (2004) [arXiv:astro-ph/0303591]; L.~Kofman, ``Probing
string theory with modulated cosmological fluctuations,''
arXiv:astro-ph/0303614; A.~Mazumdar and M.~Postma,
  ``Evolution of primordial perturbations and a fluctuating decay 
rate,''
   Phys.\ Lett.\ B {\bf 573}, 5 (2003)
   [Erratum-ibid.\ B {\bf 585}, 295 (2004)]
   [arXiv:astro-ph/0306509]; F.~Bernardeau, L.~Kofman and J.~P.~Uzan, ``Modulated
fluctuations from hybrid inflation,''  Phys.\ Rev.\ D {\bf 70}, 083004
(2004)  [arXiv:astro-ph/0403315];  D.~H.~Lyth,
``Generating the curvature perturbation at the end of inflation,''
  JCAP {\bf 0511}, 006 (2005)
  [arXiv:astro-ph/0510443].   
  
 
\bibitem{Wands:2007bd}
  D.~Wands,
``Multiple field inflation,''
  arXiv:astro-ph/0702187.
   
   
  \bibitem{StLin}
P.~J.~Steinhardt, ``Natural Inflation,'' In: {\it  The Very Early
Universe}, ed. G.W. Gibbons, S.W. Hawking and S.Siklos, Cambridge
University Press, (1983).

  \bibitem{linde1982}
A.D. Linde,
``Nonsingular Regenerating Inflationary Universe,''
Print-82-0554, Cambridge University preprint, 1982,
see http://www.stanford.edu/$\sim$alinde/1982.pdf.

\bibitem{Vilenkin:xq} A.~Vilenkin, ``The Birth Of Inflationary
Universes,'' Phys.\ Rev.\ D {\bf 27}, 2848 (1983).




\bibitem{Eternal}
A.~D.~Linde, ``Eternally Existing Self-reproducing Chaotic
Inflationary Universe,'' Phys.\ Lett.\ B {\bf 175}, 395 (1986); A.~S.~Goncharov, A.~D.~Linde and V.~F.~Mukhanov,
``The Global Structure Of The Inflationary Universe,''
  Int.\ J.\ Mod.\ Phys.\ A {\bf 2}, 561 (1987).
  
  

\bibitem{LLM}
A.~D.~Linde, D.~A.~Linde and A.~Mezhlumian, ``From the Big Bang
theory to the theory of a stationary universe,'' Phys.\ Rev.\ D
{\bf 49}, 1783 (1994) [arXiv:gr-qc/9306035].

\bibitem{nuff}
A.D. Linde,
``The New Inflationary Universe Scenario,''
In: {\it   The Very Early Universe}, ed. G.W. Gibbons, S.W. Hawking and S.Siklos,
Cambridge University Press (1983),   pp. 205-249, see  http://www.stanford.edu/$\sim$alinde/1983.pdf.


\bibitem{Borde:2001nh}
  A.~Borde, A.~H.~Guth and A.~Vilenkin,
  ``Inflationary space-times are incomplete in past directions,''
  Phys.\ Rev.\ Lett.\  {\bf 90}, 151301 (2003)
  [arXiv:gr-qc/0110012].



\bibitem{TopInf}
  A.~D.~Linde,
  ``Monopoles as big as a universe,''
  Phys.\ Lett.\ B {\bf 327}, 208 (1994)
  [arXiv:astro-ph/9402031]; A.~Vilenkin,
``Topological inflation,''
  Phys.\ Rev.\ Lett.\  {\bf 72}, 3137 (1994)
  [arXiv:hep-th/9402085]; A.~D.~Linde and D.~A.~Linde,
  ``Topological defects as seeds for eternal inflation,''
  Phys.\ Rev.\ D {\bf 50}, 2456 (1994)
  [arXiv:hep-th/9402115]; I.~Cho and A.~Vilenkin,
``Spacetime structure of an inflating global monopole,''
  Phys.\ Rev.\ D {\bf 56}, 7621 (1997)
  [arXiv:gr-qc/9708005].
  
\bibitem{Aryal:1987vn}
  M.~Aryal and A.~Vilenkin,
``The Fractal Dimension Of Inflationary Universe,''
  Phys.\ Lett.\ B {\bf 199}, 351 (1987).
  




\bibitem{Dodelson:2003ip}
S.~Dodelson, ``Coherent phase argument for inflation,'' AIP Conf.\
Proc.\  {\bf 689}, 184 (2003) [arXiv:hep-ph/0309057].

\bibitem{Mukhanov:2003xr}
 V.~F.~Mukhanov,
``CMB-slow, or How to Estimate Cosmological Parameters by Hand,''
  Int.\ J.\ Theor.\ Phys.\  {\bf 43}, 623 (2004)
  [arXiv:astro-ph/0303072].
  
   
\bibitem{Efstathiou:2003wr}
G.~Efstathiou, ``The Statistical Significance of the Low CMB
Multipoles,'' arXiv:astro-ph/0306431; G.~Efstathiou,
``A Maximum Likelihood Analysis of the Low CMB Multipoles from WMAP,''
  Mon.\ Not.\ Roy.\ Astron.\ Soc.\  {\bf 348}, 885 (2004)
  [arXiv:astro-ph/0310207]. A.~Slosar and U.~Seljak,
``Assessing the effects of foregrounds and sky removal in WMAP,''
  Phys.\ Rev.\ D {\bf 70}, 083002 (2004)
  [arXiv:astro-ph/0404567].

\bibitem{Land:2006bn}
  K.~Land and J.~Magueijo,
``The Axis of Evil revisited,''
  arXiv:astro-ph/0611518.
  
\bibitem{Rakic:2007ve}
  A.~Rakic and D.~J.~Schwarz,
``Correlating anomalies of the microwave sky: The Good, the Evil and the
Axis,''
  arXiv:astro-ph/0703266.
  
  \bibitem{Contaldi}
C.~R.~Contaldi, M.~Peloso, L.~Kofman and A.~Linde, ``Suppressing
the lower Multipoles in the CMB Anisotropies,'' JCAP {\bf 0307},
002 (2003) [arXiv:astro-ph/0303636].
  
\bibitem{Stebbins:1988bs}
  A.~Stebbins and M.~S.~Turner,
 ``Is the Great Attractor Really a Great Wall?,''
  Astrophys.\ J.\  {\bf 339}, L13 (1989).
  
\bibitem{Turner:1990uw}
  M.~S.~Turner, R.~Watkins and L.~M.~Widrow,
``Microwave distortions from collapsing domain wall bubbles,''
  Astrophys.\ J.\  {\bf 367}, L43 (1991).

\bibitem{Battye:2006mb}
  R.~A.~Battye and A.~Moss,
``Anisotropic perturbations due to dark energy,''
  Phys.\ Rev.\  D {\bf 74}, 041301 (2006)
  [arXiv:astro-ph/0602377].




  \bibitem{deVega:2006un}
  H.~J.~de Vega and N.~G.~Sanchez,
``Predictions of single field inflation for the tensor/scalar ratio and the
running spectral index,''
  Phys.\ Rev.\  D {\bf 74}, 063519 (2006).
  
\bibitem{Kallosh:2007wm}
  R.~Kallosh and A.~Linde,
``Testing String Theory with CMB,''
 arXiv:arXiv:0704.0647.
  
\bibitem{Hodges:1989dw}
  H.~M.~Hodges, G.~R.~Blumenthal, L.~A.~Kofman and J.~R.~Primack,
``Nonstandard primordial fluctuations from a polynomial inflaton potential,''
  Nucl.\ Phys.\  B {\bf 335}, 197 (1990).
  
\bibitem{Destri:2007pv}
  C.~Destri, H.~J.~de Vega and N.~G.~Sanchez,
``MCMC analysis of WMAP3 data points to broken symmetry inflaton potentials
and provides a lower bound on the tensor to scalar ratio,''
  arXiv:astro-ph/0703417.


\bibitem{SperTur} D. Spergel and N. Turok, ``Textures and cosmic structure,'' Scientific
American {\bf 266}, 52 (1992).



\bibitem{PBB} G.~Veneziano,
  ``Scale Factor Duality For Classical And Quantum Strings,''
  Phys.\ Lett.\ B {\bf 265}, 287 (1991); M.~Gasperini and G.~Veneziano,
``Pre - big bang in string cosmology,''
  Astropart.\ Phys.\  {\bf 1}, 317 (1993)
  [arXiv:hep-th/9211021].

\bibitem{nogo}
N.~Kaloper, R.~Madden and K.~A.~Olive, ``Towards a singularity -
free inflationary universe?''  Nucl.\ Phys.\ B {\bf 452}, 677
(1995) [arXiv:hep-th/9506027];
N.~Kaloper, R.~Madden and K.~A.~Olive,
  ``Axions and the Graceful Exit Problem in String Cosmology,''
  Phys.\ Lett.\ B {\bf 371}, 34 (1996)
  [arXiv:hep-th/9510117].


\bibitem{Kaloper:1998eg}
  N.~Kaloper, A.~D.~Linde and R.~Bousso,
``Pre-big-bang requires the universe to be exponentially large from the  very
beginning,''
  Phys.\ Rev.\  D {\bf 59}, 043508 (1999)
  [arXiv:hep-th/9801073];
  A.~Buonanno and T.~Damour,
``The fate of classical tensor inhomogeneities in pre-big-bang string
cosmology,''
  Phys.\ Rev.\  D {\bf 64}, 043501 (2001)
  [arXiv:gr-qc/0102102].

\bibitem{KOST}
J.~Khoury, B.~A.~Ovrut, P.~J.~Steinhardt and N.~Turok, ``The
ekpyrotic universe: Colliding branes and the origin of the hot big
bang,'' Phys.\ Rev.\ D {\bf 64}, 123522 (2001)
[arXiv:hep-th/0103239].



\bibitem{KKL}
R.~Kallosh, L.~Kofman and A.~D.~Linde, ``Pyrotechnic universe,''
Phys.\ Rev.\ D {\bf 64}, 123523 (2001) [arXiv:hep-th/0104073].



\bibitem{KKLTS}
R.~Kallosh, L.~Kofman, A.~D.~Linde and A.~A.~Tseytlin, ``BPS
branes in cosmology,'' Phys.\ Rev.\ D {\bf 64}, 123524 (2001)
[arXiv:hep-th/0106241].



\bibitem{cyclic}
P.~J.~Steinhardt and N.~Turok, ``Cosmic evolution in a cyclic
universe,'' Phys.\ Rev.\ D {\bf 65}, 126003 (2002)
[arXiv:hep-th/0111098].

\bibitem{Felder:2002jk}
  G.~N.~Felder, A.~V.~Frolov, L.~Kofman and A.~V.~Linde,
``Cosmology with negative potentials,''
  Phys.\ Rev.\  D {\bf 66}, 023507 (2002)
  [arXiv:hep-th/0202017].
  
\bibitem{Linde:2002ws}
  A.~Linde,
``Inflationary theory versus ekpyrotic / cyclic scenario,'' in {\it The future of theoretical physics and cosmology}, (Cambridge Univ. Press, Cambridge 2002), p. 801
  [arXiv:hep-th/0205259].
  


\bibitem{Liu:2002ft}
H.~Liu, G.~Moore and N.~Seiberg, ``Strings in a time-dependent
orbifold,'' JHEP {\bf 0206}, 045 (2002) [arXiv:hep-th/0204168].

\bibitem{Horowitz:2002mw}
G.~T.~Horowitz and J.~Polchinski, ``Instability of spacelike and
null orbifold singularities,'' Phys.\ Rev.\ D {\bf 66}, 103512
(2002) [arXiv:hep-th/0206228].

\bibitem{Berkooz}  M. Berkooz and  D. Reichmann, ``A Short Review of Time Dependent Solutions and Space-like Singularities in String Theory,'' arXiv:0705.2146 [hep-th].

  
\bibitem{Koyama:2007mg}
  K.~Koyama and D.~Wands,
``Ekpyrotic collapse with multiple fields,''
  arXiv:hep-th/0703040;
  K.~Koyama, S.~Mizuno, and D.~ Wands, ``Curvature perturbations from ekpyrotic collapse with multiple fields,'' arXiv:0704.1152.
  
  \bibitem{Sasaki} M. Sasaki, private communication.
  
  
  
\bibitem{Erickson:2006wc}
  J.~K.~Erickson, S.~Gratton, P.~J.~Steinhardt and N.~Turok,
``Cosmic perturbations through the cyclic ages,''
  arXiv:hep-th/0607164.


  
\bibitem{Arkani-Hamed:2003uy}
  N.~Arkani-Hamed, H.~C.~Cheng, M.~A.~Luty and S.~Mukohyama,
``Ghost condensation and a consistent infrared modification of gravity,''
  JHEP {\bf 0405}, 074 (2004)
  [arXiv:hep-th/0312099];
    N.~Arkani-Hamed, H.~C.~Cheng, M.~A.~Luty, S.~Mukohyama and T.~Wiseman,
``Dynamics of gravity in a Higgs phase,''
  JHEP {\bf 0701}, 036 (2007)
  [arXiv:hep-ph/0507120].
  
  
\bibitem{Creminelli:2006xe}
  P.~Creminelli, M.~A.~Luty, A.~Nicolis and L.~Senatore,
   ``Starting the universe: Stable violation of the null energy condition and
non-standard cosmologies,''
  JHEP {\bf 0612}, 080 (2006)
  [arXiv:hep-th/0606090].
  

\bibitem{Buchbinder:2007ad}
  E.~I.~Buchbinder, J.~Khoury and B.~A.~Ovrut,
``New ekpyrotic cosmology,''
  arXiv:hep-th/0702154.
  
\bibitem{Creminelli:2007aq}
  P.~Creminelli and L.~Senatore,
  ``A smooth bouncing cosmology with scale invariant spectrum,''
  arXiv:hep-th/0702165.
  
  
\bibitem{Adams:2006sv}
  A.~Adams, N.~Arkani-Hamed, S.~Dubovsky, A.~Nicolis and R.~Rattazzi,
``Causality, analyticity and an IR obstruction to UV completion,''
  JHEP {\bf 0610}, 014 (2006)
  [arXiv:hep-th/0602178].
  
\bibitem{Arkani-Hamed:2007ky}
  N.~Arkani-Hamed, S.~Dubovsky, A.~Nicolis, E.~Trincherini and G.~Villadoro,
``A Measure of de Sitter Entropy and Eternal Inflation,''
  arXiv:0704.1814 [hep-th].
  
\bibitem{Lehners:2007ac}
  J.~L.~Lehners, P.~McFadden, N.~Turok and P.~J.~Steinhardt,
``Generating ekpyrotic curvature perturbations before the big bang,''
  arXiv:hep-th/0702153;
   A.~J.~Tolley and D.~H.~Wesley,
``Scale-invariance in expanding and contracting universes from two-field
models,''
  arXiv:hep-th/0703101.
  
  
  
  
\bibitem{Nayeri:2005ck}
  A.~Nayeri, R.~H.~Brandenberger and C.~Vafa,
``Producing a scale-invariant spectrum of perturbations in a Hagedorn  phase
of string cosmology,''
  Phys.\ Rev.\ Lett.\  {\bf 97}, 021302 (2006)
  [arXiv:hep-th/0511140].
  
\bibitem{Brandenberger:2006xi}
  R.~H.~Brandenberger, A.~Nayeri, S.~P.~Patil and C.~Vafa,
``Tensor modes from a primordial Hagedorn phase of string cosmology,''
  arXiv:hep-th/0604126.
  
  
\bibitem{Kaloper:2006xw}
  N.~Kaloper, L.~Kofman, A.~Linde and V.~Mukhanov,
``On the new string theory inspired mechanism of generation of  cosmological
perturbations,''
  JCAP {\bf 0610}, 006 (2006)
  [arXiv:hep-th/0608200].
  

  
\bibitem{Nayeri:2006uy}
  A.~Nayeri,
``Inflation free, stringy generation of scale-invariant cosmological
  fluctuations in D = 3+1 dimensions,''
  arXiv:hep-th/0607073.
  
\bibitem{Brandenberger:2006vv}
  R.~H.~Brandenberger, A.~Nayeri, S.~P.~Patil and C.~Vafa,
``String gas cosmology and structure formation,''
  arXiv:hep-th/0608121.
  


\bibitem{Brandenberger:2006pr}
  R.~H.~Brandenberger {\it et al.},
``More on the spectrum of perturbations in string gas cosmology,''
  JCAP {\bf 0611}, 009 (2006)
  [arXiv:hep-th/0608186].
  
\bibitem{Biswas:2006bs}
  T.~Biswas, R.~Brandenberger, A.~Mazumdar and W.~Siegel,
  ``Non-perturbative gravity, Hagedorn bounce and CMB,''
  arXiv:hep-th/0610274.
   
\bibitem{Brandenberger:2007qi}
  R.~H.~Brandenberger,
``Conceptual problems of inflationary cosmology and a new approach to
cosmological structure formation,''
  arXiv:hep-th/0701111.
  
\bibitem{Germani:2006pf}
  C.~Germani, N.~E.~Grandi and A.~Kehagias,
``A stringy alternative to inflation: The cosmological slingshot scenario,''
  arXiv:hep-th/0611246.
  
\bibitem{Kachru:2002kx}
  S.~Kachru and L.~McAllister,
``Bouncing brane cosmologies from warped string compactifications,''
  JHEP {\bf 0303}, 018 (2003)
  [arXiv:hep-th/0205209].
  
  \bibitem{Hollands:2002yb}
  S.~Hollands and R.~M.~Wald,
``An alternative to inflation,''
  Gen.\ Rel.\ Grav.\  {\bf 34}, 2043 (2002)
  [arXiv:gr-qc/0205058].
  

  
\bibitem{Peter:2006hx}
  P.~Peter, E.~J.~C.~Pinho and N.~Pinto-Neto,
``A non inflationary model with scale invariant cosmological perturbations,''
  Phys.\ Rev.\  D {\bf 75}, 023516 (2007)
  [arXiv:hep-th/0610205].
  
  \bibitem{Peterprivite}  P.~Peter, E.~J.~C.~Pinho and N.~Pinto-Neto, private communication.
  
  \bibitem{LythRiotto}
D.~H.~Lyth and A.~Riotto, ``Particle physics models of inflation
and the cosmological density  perturbation,'' Phys.\ Rept.\  {\bf
314}, 1 (1999) [hep-ph/9807278].


  


\bibitem{Kallosh:1995hi}
R.~Kallosh, A.~Linde, D.~Linde and L.~Susskind, ``Gravity and
global symmetries,'' Phys.\ Rev.\  {\bf D52}, 912 (1995)
[hep-th/9502069].

  \bibitem{300} A.D. Linde, ``Inflation And Quantum Cosmology,'' Print-
86-0888 (June 1986), in {\it 300 Years of Gravitation}, ed. by S.W. Hawking and W. Israel,
Cambridge University Press, Cambridge (1987); J.~Garriga, A.~Linde and A.~Vilenkin,
``Dark energy equation of state and anthropic selection,''
  Phys.\ Rev.\  D {\bf 69}, 063521 (2004)
  [arXiv:hep-th/0310034].




\bibitem{F}
E.~J.~Copeland, A.~R.~Liddle, D.~H.~Lyth, E.~D.~Stewart and
D.~Wands, ``False vacuum inflation with Einstein gravity,'' Phys.\
Rev.\ D {\bf 49}, 6410 (1994) [astro-ph/9401011]; G.~R.~Dvali,
Q.~Shafi and R.~Schaefer, ``Large scale structure and
supersymmetric inflation without fine tuning,'' Phys.\ Rev.\
Lett.\  {\bf 73}, 1886 (1994) [hep-ph/9406319];
 A.~D.~Linde and A.~Riotto,
``Hybrid inflation in supergravity,'' Phys.\ Rev.\ D {\bf 56},
1841 (1997) [arXiv:hep-ph/9703209].


\bibitem{D}
P.~Binetruy and G.~Dvali, ``D-term inflation,'' Phys.\ Lett.\ B
{\bf 388}, 241 (1996) [hep-ph/9606342]; E.~Halyo, ``Hybrid
inflation from supergravity D-terms,'' Phys.\ Lett.\ B {\bf 387},
43 (1996) [hep-ph/9606423].




\bibitem{pterm}
R.~Kallosh and A.~Linde, ``P-term, D-term and F-term inflation,''
JCAP {\bf 0310}, 008 (2003) [arXiv:hep-th/0306058].

\bibitem{Binetruy:2004hh}
  P.~Binetruy, G.~Dvali, R.~Kallosh and A.~Van Proeyen,
``Fayet-Iliopoulos terms in supergravity and cosmology,''
  Class.\ Quant.\ Grav.\  {\bf 21}, 3137 (2004)
  [arXiv:hep-th/0402046].
  
\bibitem{Lyth:2007qh}
  D.~H.~Lyth,
``Particle physics models of inflation,''
  arXiv:hep-th/0702128.
  

  
\bibitem{Kallosh:2007ig}
  R.~Kallosh,
``On inflation in string theory,''
  arXiv:hep-th/0702059.
  


\bibitem{jap}
M.~Kawasaki, M.~Yamaguchi and T.~Yanagida, ``Natural chaotic
inflation in supergravity,'' Phys.\ Rev.\ Lett.\  {\bf 85}, 3572
(2000) [arXiv:hep-ph/0004243].

\bibitem{Yamaguchi:2001pw}
M.~Yamaguchi and J.~Yokoyama, ``New inflation in supergravity with
a chaotic initial condition,'' Phys.\ Rev.\ D {\bf 63}, 043506
(2001) [arXiv:hep-ph/0007021]; M.~Yamaguchi, ``Natural double
inflation in supergravity,'' Phys.\ Rev.\ D {\bf 64}, 063502
(2001) [arXiv:hep-ph/0103045].


\bibitem{Yok}
M.~Yamaguchi and J.~Yokoyama,
``Chaotic hybrid new inflation in supergravity with a running spectral
index,''
  Phys.\ Rev.\  D {\bf 68}, 123520 (2003)
  [arXiv:hep-ph/0307373].
  
  \bibitem{natural}
K.~Freese, J.~A.~Frieman and A.~V.~Olinto,
 ``Natural Inflation With Pseudo - Nambu-Goldstone Bosons,''
Phys.\ Rev.\ Lett.\  {\bf 65}, 3233 (1990).


  
  \bibitem{Savage:2006tr}
  C.~Savage, K.~Freese and W.~H.~Kinney,
``Natural inflation: Status after WMAP 3-year data,''
  Phys.\ Rev.\  D {\bf 74}, 123511 (2006)
  [arXiv:hep-ph/0609144].
  
  
\bibitem{GKP}
S.~B.~Giddings, S.~Kachru and J.~Polchinski, ``Hierarchies from
fluxes in string compactifications,'' Phys. Rev. {\bf D66}, 106006
(2002) [arXiv:hep-th/0105097].


\bibitem{KKLT}
S.~Kachru, R.~Kallosh, A.~Linde and S.~P.~Trivedi, ``De Sitter
vacua in string theory,'' Phys.\ Rev.\ D {\bf 68}, 046005 (2003)
[arXiv:hep-th/0301240].


\bibitem{Burgess:2003ic}
C.~P.~Burgess, R.~Kallosh and F.~Quevedo, ``de Sitter string vacua
from supersymmetric D-terms,'' JHEP {\bf 0310}, 056 (2003)
[arXiv:hep-th/0309187].

\bibitem{Str} E.~Silverstein,
``(A)dS backgrounds from asymmetric orientifolds,''
arXiv:hep-th/0106209;
A.~Maloney, E.~Silverstein and A.~Strominger, ``De Sitter space in
noncritical string theory,'' arXiv:hep-th/0205316.



\bibitem{Arkani-Hamed:1998rs}
  N.~Arkani-Hamed, S.~Dimopoulos and G.~R.~Dvali,
``The hierarchy problem and new dimensions at a millimeter,''
  Phys.\ Lett.\ B {\bf 429}, 263 (1998)
  [arXiv:hep-ph/9803315]; I.~Antoniadis, N.~Arkani-Hamed, S.~Dimopoulos and G.~R.~Dvali,
``New dimensions at a millimeter to a Fermi and superstrings at a TeV,''
  Phys.\ Lett.\ B {\bf 436}, 257 (1998)
  [arXiv:hep-ph/9804398].
  
  
 \bibitem{Randall:1999ee} L.~Randall and R.~Sundrum,
``A large mass hierarchy from a small extra dimension,''
  Phys.\ Rev.\ Lett.\  {\bf 83}, 3370 (1999)
  [arXiv:hep-ph/9905221].
  
\bibitem{Blanco-Pillado:2004ns}
  J.J. Blanco-Pillado, C.P. Burgess, J.M. Cline, C. Escoda, M. Gomez-Reino, R. Kallosh, A. Linde, F. Quevedo, ``Racetrack inflation,''
  JHEP {\bf 0411}, 063 (2004)
  [arXiv:hep-th/0406230].
  
  \bibitem{Conlon:2005jm}
  J.~P.~Conlon and F.~Quevedo,
``Kaehler moduli inflation,''
  JHEP {\bf 0601}, 146 (2006)
  [arXiv:hep-th/0509012].
  
\bibitem{Lalak:2005hr}
  Z.~Lalak, G.~G.~Ross and S.~Sarkar,
``Racetrack inflation and assisted moduli stabilisation,''
  Nucl.\ Phys.\  B {\bf 766}, 1 (2007)
  [arXiv:hep-th/0503178].
  
  
\bibitem{Blanco-Pillado:2006he}
J.J. Blanco-Pillado, C.P. Burgess, J.M. Cline, C. Escoda, M. Gomez-Reino, R. Kallosh, A. Linde, F. Quevedo,,
``Inflating in a better racetrack,''
  JHEP {\bf 0609}, 002 (2006)
  [arXiv:hep-th/0603129].

\bibitem{Bond:2006nc}
  J.~R.~Bond, L.~Kofman, S.~Prokushkin and P.~M.~Vaudrevange,
 ``Roulette inflation with Kaehler moduli and their axions,''
  arXiv:hep-th/0612197.



\bibitem{Dvali:1998pa}
G.~R.~Dvali and S.~H.~H.~Tye, ``Brane inflation,'' Phys.\ Lett.\ B
{\bf 450}, 72 (1999) [arXiv:hep-ph/9812483].


\bibitem{Quevedo}
F.~Quevedo,
``Lectures on string / brane cosmology,''
  Class.\ Quant.\ Grav.\  {\bf 19}, 5721 (2002)
  [arXiv:hep-th/0210292].



\bibitem{KKLMMT}
S.~Kachru, R.~Kallosh, A.~Linde, J.~Maldacena, L.~McAllister and
S.~P.~Trivedi, ``Towards inflation in string theory,'' JCAP {\bf
0310}, 013 (2003) [arXiv:hep-th/0308055].


  
\bibitem{Hsu:2003cy}
J.~P.~Hsu, R.~Kallosh and S.~Prokushkin, ``On brane inflation with
volume stabilization,'' JCAP {\bf 0312}, 009 (2003)
[arXiv:hep-th/0311077].



\bibitem{renata}
R.~Kallosh, ``N = 2 supersymmetry and de Sitter space,''
arXiv:hep-th/0109168; C.~Herdeiro, S.~Hirano and R.~Kallosh,
``String theory and hybrid inflation/acceleration,'' JHEP {\bf
0112} (2001) 027 [arXiv:hep-th/0110271]; K.~Dasgupta, C.~Herdeiro,
S.~Hirano and R.~Kallosh, ``D3/D7 inflationary model and
M-theory,'' Phys.\ Rev.\ D {\bf 65}, 126002 (2002)
[arXiv:hep-th/0203019];
 J.~P.~Hsu and R.~Kallosh,
``Volume stabilization and the origin of the inflaton shift symmetry in string theory,''
  JHEP {\bf 0404}, 042 (2004)
  [arXiv:hep-th/0402047].



   
  \bibitem{Dasgupta:2004dw}
  K.~Dasgupta, J.~P.~Hsu, R.~Kallosh, A.~Linde and M.~Zagermann,
``D3/D7 brane inflation and semilocal strings,''
  JHEP {\bf 0408}, 030 (2004)
  [arXiv:hep-th/0405247].

  
  \bibitem{Chen:2005ae}
  P.~Chen, K.~Dasgupta, K.~Narayan, M.~Shmakova and M.~Zagermann,
``Brane inflation, solitons and cosmological solutions: I,''
  JHEP {\bf 0509}, 009 (2005)
  [arXiv:hep-th/0501185].
     
\bibitem{Buchbinder:2004nt}
  E.~I.~Buchbinder,
``Five-brane dynamics and inflation in heterotic M-theory,''
  Nucl.\ Phys.\ B {\bf 711}, 314 (2005)
  [arXiv:hep-th/0411062].

  
  \bibitem{Becker:2005sg}
  K.~Becker, M.~Becker and A.~Krause,
``M-theory inflation from multi M5-brane dynamics,''
  arXiv:hep-th/0501130.



\bibitem{kinfl}
C.~Armendariz-Picon, T.~Damour and V.~Mukhanov, ``k-inflation,''
Phys.\ Lett.\ B {\bf 458}, 209 (1999) [arXiv:hep-th/9904075].


\bibitem{Dim}
S.~Dimopoulos and S.~Thomas, ``Discretuum versus continuum dark
energy,'' Phys.\ Lett.\ B {\bf 573}, 13 (2003)
[arXiv:hep-th/0307004].

\bibitem{Silverstein:2003hf}
E.~Silverstein and D.~Tong,
``Scalar speed limits and cosmology: Acceleration from D-cceleration,''
  Phys.\ Rev.\ D {\bf 70}, 103505 (2004)
  [arXiv:hep-th/0310221];  M.~Alishahiha, E.~Silverstein and D.~Tong,
``DBI in the sky,''
  Phys.\ Rev.\ D {\bf 70}, 123505 (2004)
  [arXiv:hep-th/0404084].  
  
\bibitem{Chen:2006hs}
  X.~Chen, S.~Sarangi, S.~H.~Henry Tye and J.~Xu,
``Is brane inflation eternal?,''
  JCAP {\bf 0611}, 015 (2006)
  [arXiv:hep-th/0608082].


\bibitem{Kallosh:2004yh}
  R.~Kallosh and A.~Linde,
``Landscape, the scale of SUSY breaking, and inflation,''
  JHEP {\bf 0412}, 004 (2004)
  [arXiv:hep-th/0411011];
  J.~J.~Blanco-Pillado, R.~Kallosh and A.~Linde,
``Supersymmetry and stability of flux vacua,''
  JHEP {\bf 0605}, 053 (2006)
  [arXiv:hep-th/0511042].
  
    \bibitem{kl2007} R. Kallosh and A. Linde, ``Testing String Theory with CMB,'' JCAP{\bf 04}, 017 (2007) [arXiv:0704.0647].


  


  
\bibitem{DeWolfe:2002nn}
O.~DeWolfe and S.~B.~Giddings,
``Scales and hierarchies in warped compactifications and brane worlds,''
Phys.\ Rev.\ D {\bf 67}, 066008 (2003)
[arXiv:hep-th/0208123].

\bibitem{Arkani-Hamed:2004fb}
N.~Arkani-Hamed and S.~Dimopoulos,
``Supersymmetric unification without low energy supersymmetry and signatures
for fine-tuning at the LHC,''
arXiv:hep-th/0405159; N.~Arkani-Hamed, S.~Dimopoulos, G.~F.~Giudice and A.~Romanino,
``Aspects of split supersymmetry,''
  Nucl.\ Phys.\ B {\bf 709}, 3 (2005)
  [arXiv:hep-ph/0409232].

\bibitem{Baumann:2006cd}
  D.~Baumann and L.~McAllister,
``A microscopic limit on gravitational waves from D-brane inflation,''
  arXiv:hep-th/0610285.

\bibitem{Bean:2007hc}
  R.~Bean, S.~E.~Shandera, S.~H.~Henry Tye and J.~Xu,
``Comparing brane inflation to WMAP,''
  arXiv:hep-th/0702107.


\bibitem{ZelStar}
Y.~B.~Zeldovich and A.~A.~Starobinsky, ``Quantum Creation Of A Universe In A
Nontrivial Topology,'' Sov.\ Astron.\ Lett.\  {\bf 10},  135 (1984).


\bibitem{chaotmix} O. Heckmann \& E. Schucking, in {\em Handbuch der Physik},
ed. S. Flugge (Springer, Berlin, 1959), Vol. 53, p.515; G.F. Ellis, Gen. Rel.
Grav. {\bf 2}, 7 (1971); J.R. Gott, ``Chaotic Cosmologies,'' { Mon. Not. R.
Astron. Soc.} {\bf 193}, 153 (1980);  C.N. Lockhart, B.Misra and I. Prigogine,
Phys. Rev. {\bf D25}, 921 (1982); H.V. Fagundes, Phys. Rev. Lett. {\bf 51}, 517
(1983).

\bibitem{topol4} N.~J.~Cornish, D.~N.~Spergel and G.~D.~Starkman,
``Does chaotic mixing facilitate $\Omega < 1$ inflation?'' Phys.\ Rev.\ Lett.\
{\bf 77}, 215 (1996) [arXiv:astro-ph/9601034].

\bibitem{Coule}
D.~H.~Coule and J.~Martin, ``Quantum cosmology and open universes,'' Phys.\
Rev.\ D {\bf 61}, 063501 (2000) [arXiv:gr-qc/9905056].



\bibitem{Linde:2004nz}
  A.~Linde,
``Creation of a compact topologically nontrivial inflationary universe,''
  JCAP {\bf 0410}, 004 (2004)
  [arXiv:hep-th/0408164].

 




\bibitem{Lerche:1986cx}
  W.~Lerche, D.~L\"ust and A.~N.~Schellekens,
 ``Chiral Four-Dimensional Heterotic Strings From Selfdual Lattices,''
  Nucl.\ Phys.\ B {\bf 287}, 477 (1987).


\bibitem{Bousso:2000xa}
  R.~Bousso and J.~Polchinski,
``Quantization of four-form fluxes and dynamical neutralization of the
cosmological constant,''
  JHEP {\bf 0006}, 006 (2000)
  [arXiv:hep-th/0004134].


\bibitem{Douglas}  M.~R.~Douglas,
{``The statistics of string / M theory vacua,''}
  JHEP {\bf 0305} 046  (2003)
  [arXiv:hep-th/0303194];
  F.~Denef and M.~R.~Douglas,
``Distributions of flux vacua,''
  JHEP {\bf 0405}, 072 (2004)
  [arXiv:hep-th/0404116];
  M.~R.~Douglas and S.~Kachru,
``Flux compactification,''
  arXiv:hep-th/0610102;
  F.~Denef, M.~R.~Douglas and S.~Kachru,
``Physics of string flux compactifications,''
  arXiv:hep-th/0701050.

 

\bibitem{Linde:2002gj} A.~Linde, ``Inflation, quantum cosmology and the anthropic principle,'' in {\it  Science and Ultimate Reality: From Quantum to Cosmos}, (eds. J.D. Barrow, P.C.W. Davies, \& C.L. Harper, Cambridge University Press, 2003) [arXiv:hep-th/0211048].

\bibitem{Susskind:2003kw}
  L.~Susskind,
  ``The anthropic landscape of string theory,''
  arXiv:hep-th/0302219.
  
\bibitem{Brown:1987dd}
  J.~D.~Brown and C.~Teitelboim,
``Dynamical neutralization of the cosmological constant,''
  Phys.\ Lett.\  B {\bf 195} (1987) 177;
  J.~D.~Brown and C.~Teitelboim,
``Neutralization of the Cosmological Constant by Membrane Creation,''
  Nucl.\ Phys.\  B {\bf 297}, 787 (1988).
  
\bibitem{Ceresole:2006iq}
  A.~Ceresole, G.~Dall'Agata, A.~Giryavets, R.~Kallosh and A.~Linde,
``Domain walls, near-BPS bubbles, and probabilities in the landscape,''
  Phys.\ Rev.\  D {\bf 74}, 086010 (2006)
  [arXiv:hep-th/0605266].
  
  \bibitem{Clifton:2007en}
  T.~Clifton, A.~Linde and N.~Sivanandam,
``Islands in the landscape,''
  JHEP {\bf 0702}, 024 (2007)
  [arXiv:hep-th/0701083].
  
  
\bibitem{Linde:1984ir}
  A.~D.~Linde,
``The Inflationary Universe,''
  Rept.\ Prog.\ Phys.\  {\bf 47}  925 (1984).
  
\bibitem{Hawking:2006ur}
  S.~W.~Hawking and T.~Hertog,
``Populating the landscape: A top down approach,''
  Phys.\ Rev.\  D {\bf 73}, 123527 (2006)
  [arXiv:hep-th/0602091].
  
  \bibitem{Linde:1987b}
A. D. Linde, 
``Inflation And Axion Cosmology,''
Phys.\ Lett.\ B {\bf 201}  (1988) 437.

\bibitem{Rees}
  M.~Tegmark, A.~Aguirre, M.~Rees and F.~Wilczek,
``Dimensionless constants, cosmology and other dark matters,''
  Phys.\ Rev.\ D {\bf 73} (2006) 023505
  [arXiv:astro-ph/0511774].
  
  \bibitem{Affleck:1984fy}
  I.~Affleck and M.~Dine,
``A New Mechanism For Baryogenesis,''
  Nucl.\ Phys.\ B {\bf 249} (1985) 361.
  
\bibitem{Linde:gh}
A. D. Linde,
``The New Mechanism Of Baryogenesis And The Inflationary Universe,''
Phys.\ Lett.\ B {\bf 160}  (1985) 243.

  \bibitem{ehr} P. Ehrenfest, 
  ``In what way does it become manifest in fundamental laws of physics that space has three dimensions?'' Proc. Amsterdam Acad. {\bf 20}, 200 (1917);
P. Ehrenfest, "How do the fundamental laws of physics make manifest that space has 3 dimensions?"  Annalen der Physik {\bf 61},  440 (1920).

  

 
\bibitem{Banks}  A. D. Sakharov, ``Cosmological
Transitions With A Change In Metric Signature,'' Sov. Phys. JETP {\bf 60}, 214 (1984) [Zh.
Eksp. Teor. Fiz. {\bf 87}, 375 (1984)];
   T. Banks, ``TCP, quantum gravity, the cosmological constant, and al l that ...,'' Nucl. Phys. B249, 332, (1985);  A.D. Linde, ``Inflation And Quantum Cosmology,'' Print-
86-0888 (June 1986), in {\it 300 Years of Gravitation}, ed. by S.W. Hawking and W. Israel,
Cambridge University Press, Cambridge (1987).



\bibitem{Weinberg:1987dv}
  S.~Weinberg,
``Anthropic Bound On The Cosmological Constant,''
  Phys.\ Rev.\ Lett.\  {\bf 59} (1987) 2607.
  
  \bibitem{Weinberg1998}
  H.~Martel, P.~R.~Shapiro and S.~Weinberg,
``Likely Values of the Cosmological Constant,''
  Astrophys.\ J.\  {\bf 492}, 29 (1998)
  [arXiv:astro-ph/9701099].
  
\bibitem{Garriga:1999hu}
  J.~Garriga, M.~Livio and A.~Vilenkin,
``The cosmological constant and the time of its dominance,''
  Phys.\ Rev.\ D {\bf 61}, 023503 (2000)
  [arXiv:astro-ph/9906210];  J.~Garriga and A.~Vilenkin,
``Solutions to the cosmological constant problems,''
  Phys.\ Rev.\ D {\bf 64}, 023517 (2001)
  [arXiv:hep-th/0011262].
  
 
  
    
\bibitem{Tegmark:2005dy}
  M.~Tegmark, A.~Aguirre, M.~Rees and F.~Wilczek,
``Dimensionless constants, cosmology and other dark matters,''
  Phys.\ Rev.\ D {\bf 73}, 023505 (2006)
  [arXiv:astro-ph/0511774].
  
\bibitem{Bousso:2007kq}
  R.~Bousso, R.~Harnik, G.~D.~Kribs and G.~Perez,
``Predicting the Cosmological Constant from the Causal Entropic Principle,''
  arXiv:hep-th/0702115.
  
    \bibitem{Lineweaver:2007qh}
  C.~H.~Lineweaver and C.~A.~Egan,
``The Cosmic Coincidence as a Temporal Selection Effect Produced by the Age
Distribution of Terrestrial Planets in the Universe,''
  arXiv:astro-ph/0703429.
  
  
  \bibitem{Starobinsky:1986fx}
  A.~A.~Starobinsky,
``Stochastic De Sitter (Inflationary) Stage In The Early Universe,''
in: {\it Current Topics in Field Theory, Quantum Gravity and
Strings}, Lecture Notes in Physics, eds. H.J. de Vega and N. Sanchez
(Springer, Heidelberg 1986) {\bf 206}, p. 107.


\bibitem{Goncharov:1987ir}
  A.~S.~Goncharov, A.~D.~Linde and V.~F.~Mukhanov,
``The Global Structure Of The Inflationary Universe,''
  Int.\ J.\ Mod.\ Phys.\ A {\bf 2}, 561 (1987).
  
\bibitem{Garriga:1997ef}
  J.~Garriga and A.~Vilenkin,
``Recycling universe,''
  Phys.\ Rev.\ D {\bf 57}, 2230 (1998)
  [arXiv:astro-ph/9707292];
  V.~Vanchurin and A.~Vilenkin,
``Eternal observers and bubble abundances in the landscape,''
  Phys.\ Rev.\  D {\bf 74}, 043520 (2006)
  [arXiv:hep-th/0605015].


 
\bibitem{Bousso:2006ev}
  R.~Bousso,
``Holographic probabilities in eternal inflation,''
  Phys.\ Rev.\ Lett.\  {\bf 97}, 191302 (2006)
  [arXiv:hep-th/0605263].
  
\bibitem{Podolsky:2007vg}
  D.~Podolsky and K.~Enqvist,
``Eternal inflation and localization on the landscape,''
  arXiv:0704.0144 [hep-th].
  
 

  

   \bibitem{Bellido} J.~Garcia-Bellido, A.~D.~Linde and D.~A.~Linde, ``Fluctuations of the gravitational constant in the inflationary Brans-Dicke cosmology,''
Phys.\ Rev.\ D {\bf 50}, 730 (1994)
  [arXiv:astro-ph/9312039];  J.~Garcia-Bellido and A.~D.~Linde,
``Stationarity of inflation and predictions of quantum cosmology,''
  Phys.\ Rev.\ D {\bf 51}, 429 (1995)
  [arXiv:hep-th/9408023];
  J.~Garcia-Bellido and A.~D.~Linde,
``Stationary solutions in Brans-Dicke stochastic inflationary cosmology,''
  Phys.\ Rev.\ D {\bf 52}, 6730 (1995)
  [arXiv:gr-qc/9504022].



\bibitem{Mediocr}
  A.~Vilenkin,
``Predictions from quantum cosmology,''
  Phys.\ Rev.\ Lett.\  {\bf 74}, 846 (1995)
  [arXiv:gr-qc/9406010].
  
 
 


  \bibitem{Vilenkin:2006qf}
A.~Vilenkin, ``Making predictions in eternally inflating universe,''
  Phys.\ Rev.\ D {\bf 52}, 3365 (1995)
  [arXiv:gr-qc/9505031];  S.~Winitzki and A.~Vilenkin,
``Uncertainties of predictions in models of eternal inflation,''
  Phys.\ Rev.\ D {\bf 53}, 4298 (1996)
  [arXiv:gr-qc/9510054];  
  A.~Vilenkin,
``Unambiguous probabilities in an eternally inflating universe,''
  Phys.\ Rev.\ Lett.\  {\bf 81}, 5501 (1998)
  [arXiv:hep-th/9806185];  V.~Vanchurin, A.~Vilenkin and S.~Winitzki,
``Predictability crisis in inflationary cosmology and its resolution,''
  Phys.\ Rev.\  D {\bf 61}, 083507 (2000)
  [arXiv:gr-qc/9905097]; J.~Garriga and A.~Vilenkin,
``A prescription for probabilities in eternal inflation,''
  Phys.\ Rev.\ D {\bf 64}, 023507 (2001)
  [arXiv:gr-qc/0102090];     R.~Easther, E.~A.~Lim and M.~R.~Martin,
``Counting pockets with world lines in eternal inflation,''
  JCAP {\bf 0603}, 016 (2006)
  [arXiv:astro-ph/0511233]; A.~Vilenkin,
  ``Probabilities in the landscape,''
  arXiv:hep-th/0602264;
    V.~Vanchurin,
``Geodesic Measures of the Landscape,''
  arXiv:hep-th/0612215.
  
\bibitem{GDVW}  J.~Garriga, D.~Schwartz-Perlov, A.~Vilenkin and S.~Winitzki,
  ``Probabilities in the inflationary multiverse,''
  JCAP {\bf 0601}, 017 (2006)
  [arXiv:hep-th/0509184]; 

 
\bibitem{Vilenkin:2006qg}
  A.~Vilenkin,
``Freak observers and the measure of the multiverse,''
  JHEP {\bf 0701}, 092 (2007)
  [arXiv:hep-th/0611271].
  
   \bibitem{volume}  {This family of probability measures sometimes are called ``global,'' whereas the measures based on the comoving coordinates are called ``local,'' see e.g. \cite{Bousso:2006xc}.  However, this terminology, and the often repeated implication that global means acausal,  are somewhat misleading. All of these measures are based on investigation of the physical evolution of  a single causally connected domain of initial size $O(H^{-1})$. The early stages of the evolution of our part of the universe were influenced by the evolution of other parts of the original domain even though some of these parts at present are exponentially far away from us. One should not confuse the exponentially large particle horizon, which is relevant for understanding of the origin of our part of the universe, with the present radius of the event horizon $H^{{-1}} \sim 10^{28}$ cm, which is relevant for understanding of our future.} 
   
\bibitem{Bousso:2006xc}
  R.~Bousso and B.~Freivogel,
 ``A paradox in the global description of the multiverse,''
  arXiv:hep-th/0610132.

\bibitem{Guth:2007ng}
A.~H.~Guth,
``Inflation,''
  arXiv:astro-ph/0404546.
  A.~H.~Guth,
``Eternal inflation and its implications,''
  arXiv:hep-th/0702178.
  
\bibitem{Tegmark:2004qd}
 M.~Tegmark,
 ``What does inflation really predict?,''
  JCAP {\bf 0504}, 001 (2005)
  [arXiv:astro-ph/0410281].
  
  

  
  
  
  \bibitem{lindemeasure} 
  A.~Linde,
``Towards a gauge invariant volume-weighted probability measure for eternal inflation,''
  arXiv:0705.1160 [hep-th].
  
\bibitem{Aguirre:2006ak}
  A.~Aguirre, S.~Gratton and M.~C.~Johnson,
``Hurdles for recent measures in eternal inflation,''
  arXiv:hep-th/0611221.
     
  
\bibitem{Winitzki:2006rn}
  S.~Winitzki,
``Predictions in eternal inflation,''
  arXiv:gr-qc/0612164.

  
\bibitem{Garriga:2005ee}
  J.~Garriga and A.~Vilenkin,
``Anthropic prediction for Lambda and the Q catastrophe,''
  Prog.\ Theor.\ Phys.\ Suppl.\  {\bf 163}, 245 (2006)
  [arXiv:hep-th/0508005].
  
  
  \bibitem{Harnik:2006vj}
  R.~Harnik, G.~D.~Kribs and G.~Perez,
``A universe without weak interactions,''
  Phys.\ Rev.\ D {\bf 74}, 035006 (2006)
  [arXiv:hep-ph/0604027].




  \bibitem{Starkman:2006at}
  G.~D.~Starkman and R.~Trotta,
``Why anthropic reasoning cannot predict Lambda,''
  Phys.\ Rev.\ Lett.\  {\bf 97}, 201301 (2006)
  [arXiv:astro-ph/0607227].
  

  
  
\end{thebibliography}
\end{document}